\documentclass[11pt,a4paper]{article}
\usepackage{epsfig,graphicx,amsfonts,amsmath}

\usepackage{cite}
\RequirePackage{mathrsfs}
\usepackage{color}
\usepackage{epsf}
\usepackage{epstopdf}

\newlength{\dinwidth}
\newlength{\dinmargin}
\def\eq#1{{Eq.~(\ref{#1})}}
\newcommand{\Le}{\left(}
\newcommand{\Ra}{\right)}

\newcommand{\beq}{\begin{equation}}
\newcommand{\eeq}{\end{equation}}
\newcommand{\beqar}{\begin{eqnarray}}
\newcommand{\eeqar}{\end{eqnarray}}
\newcommand{\D}{\partial}
\newcommand{\g}{{\rm g}}

\newcommand{\ep}{\varepsilon}

\newcommand{\tv}{\textsl{v}}

\newcommand{\B}{{\cal B}}
\newcommand{\T}{{\cal T}}
\newcommand{\overl}{\overline}
\newcommand\cev[1]{\overleftarrow{#1}}

\date{}
\begin{document}

\title {{~}\\
{\Large \bf Unifying approaches: derivation of Balitsky hierarchy from the Lipatov effective action }}

\author{ 
{~}\\
{\large 
S.~Bondarenko$^{(1) }$,
S.~Pozdnyakov$^{(2) }$,
A.~Prygarin$^{(1) }$
}\\[7mm]
{\it\normalsize  $^{(1) }$ Physics Department, Ariel University, Ariel 40700, Israel}\\
{\it\normalsize  $^{(2) }$ St.Petersburg State University, St. Petersburg 199034, Russia}\\
}

\maketitle
\thispagestyle{empty}

\begin{abstract}
 We consider a derivation of the hierarchy of correlators of ordered exponentials directly from the Lipatov's effective action~\cite{LipatovEff}
formulated in terms of  interacting ordered exponentials~\cite{OurZub}.
The  derivation of the Balitsky equation~\cite{Bal} from the hierarchy is discussed as well as
the way the sub-leading eikonal corrections to the Balitsky equation arise from the transverse field contribution  and sub-leading eikonal corrections to the quark propagator. 
We outline other possible applications of the proposed calculation scheme.
\end{abstract}

\section{Introduction}

The Lipatov's effective action approach~\cite{LipatovEff} and the
 Balitsky-Kovchegov~(BK)- Jalilian-Marian, Iancu, McLerran, Weigert, Leonidov and Kovner (JIMWLK)
 approaches~\cite{Bal,Kovch,Venug,JIMWLK}, see also examples of the phenomenological applications in ~\cite{BKexp}, are
  both 
 intended to describe amplitudes of QCD scattering
at high-energies, 
despite the fact that the two independent  approaches 
are formulated in terms of different degrees of freedom.
The Lipatov's action 
deals 
with scattering 
amplitudes in terms of reggeized gluons, it allows to calculate
quasi-elastic amplitudes of high-energy scattering processes
in the  multi-Regge kinematics. There are many applications of this action for the description of  high energy processes
and calculation of sub-leading unitarizing  corrections to the
amplitudes and production vertices,
 see for example  \cite{Our1,Our2,Our3,Our4,Our5,EffAct,Fadin}. 
In the BK-JIMWLK approach on the other hand~\cite{Bal,Kovch,Venug,JIMWLK} 
one treats the scattering problems in terms of correlators of the ordered exponentials written through  external  fields.

In our previous studies~\cite{Our1,Our2,Our3,Our4} we demonstrated that  the Color Glass  Condensate~(CGC) and corresponding JIMLK approaches are interconnected with the 
effective action of Lipatov. In our further studies~\cite{OurZub,Our4} we made a step towards the unification of the two  approaches, where the Lipatov's action was reformulated in terms of interacting
ordered exponentials, which are main degrees of freedom of Balitsky hierarchy of equations, \cite{Bal}.
Our ultimate goal is  to unify the approaches of Lipatov and Balitsky
to the description of the scattering at high-energies in the framework valid for both variants of the theory. In particular,  
we consider  the derivation of Balitsky hierarchy for the quark (fundamental) Wilson lines
from the action of Lipatov with the use of \cite{OurZub,Our4} results.
The similar calculations were done in \cite{Hetch}, there the comparison between the approaches was performed on the base of the diagrammatic calculations
with the use of the vertices and Feynman rules obtained from the Lipatov's, \cite{LipatovEff}, action. 
Nevertheless, in our calculations we consider an another way to derive the hierarchy of the correlators.
This derivation we implement as follows. Concerning the Lipatov's effective action,  
we can derive the action beginning from the QCD action and introducing there the interacting exponentials as  an additional part of the action, see discussion and details in \cite{OurZub},
the exponentials provide a non-local rapidity dependence in the description of the scattering. 
At the next step, the product of the exponentials is eliminated from the action by the introduction
of the auxiliary fields,
the Reggeons. In the transformed action those exponentials are represented separately as effective covariant currents with Reggeons as their sources~\cite{Our1,Our2} and
this is the original, Lipatov's one, action, see \cite{LipatovEff}.
The whole rapidity interval of a scattering, consequently,  is labeled and divided to the different QCD clusters at some values of rapidity variable, 
the interaction between the  clusters are due the Reggeons and this is a classical description of the 
high energy Balitsky-Fadin-Kuraev-Lipatov~\cite{BFKL} approach to the scattering in the Regge kinematics.
Nevertheless, we can perform the same calculations directly on the level of the complimentary variant of the Lipatov's action, \cite{OurZub},  with only ordered exponentials included considering them as the objects of interests.
In this case, applying the light-cone gauge, the gluon's propagator will not affected by the additional terms in the QCD Lagrangian 
and we obtain a theory of interacting Wilson lines averaged with respect to the gluon and quarks fields for the each separated QCD cluster.

 This formulation of the problem has a few advantages in comparison to the standard derivation of the correlators of Wilson lines.  First of all, we deal with the usual QCD Lagrangian, the introduction of the different gauges and corresponding calculations of the propagators in this case is a standard and well defined  procedure.
An important consequence of this is that it allows systematically define and calculate any kind of the many loops corrections to the correlators in correspondence to any gauge chosen in the initial action.
An another advantage of the approach is that we obtain 
the equations for the different correlators directly from the action, without the approximation which concerns the external fields. In turn, the form and structure of the external fields 
can be borrowed from the complimentary Lipatov's action written in terms of reggeized gluons. The external fields there are 
reggeized gluons as sources of the effective currents, see details in \cite{Our1,Our2} and this is the way how the reggeized gluon fields are entering in the hierarchy of equations of Balitsky.
The shock wave approximation to the fields, required for the derivation of the original BK equation, is applied further in this case, the initial equations for the correlator are correct beyond the shock wave approximation as mentioned above.

 Another important issue  discussed in the paper is the origin and the form of   ordered exponentials in the Lipatov's action.
In the papers ~\cite{OurZub,TwoD} it was demonstrated how exponentials are arising in the effective QCD action. Following to the \cite{Bal}, we can consider this additional part of the action as
arising from a truncated propagator of the energetic particle in an external field, see Section~\ref{SS}.
The additional corrections to the ordered exponential in this case, both sub-leading eikonal and sub-leading non-eikonal, arise in the definition of the effective currents. The usual Wilson line is a leading order~(LO) value of such object but in general we have to distinguish the two types of the Wilson lines, for the quarks (fundamental) and for the gluons (adjoint). In the action the currents 
appear after the color averaging, the form of the currents is not important consequently to the LO precision 
and it does not affects on the further calculations. Nevertheless, if we account the contribution from the truncated propagator as the operator in the effective action, the difference is important. The  propagators of quark and gluon in an external field are not the same, the sub-leading and non-eikonal corrections they provide are not the same as well. Therefore, for mostly general action, we need to introduce a two forms of the Wilson lines directly in the Lipatov's action and in it's modified form. Still, in the present article we derive the Balitsky hierarchy only for the quark Wilson lines, the same hierarchy for the adjoint Wilson lines will be different only if we will account sub-leading and non-eikonal corrections to the obtained system of the equations.

The paper is written as follows. 
In  Section~\ref{SS}   we discuss the appearance and role of the operators related to the ordered exponentials in the action. In  Section \ref{WL} we consider
the effective action of Lipatov with those operators included in the 	Lagrangian and  the derivation of general BK-like equation for the 
correlators of Wilson lines.
In Section~\ref{Ef} we discuss the gluon's propagator requested for the calculation of the correlator whereas
Section~\ref{BKE} is dedicated to the application of the shock wave approximation to the correlator's 
equation and derivation of the original form of the Balitsky's equation. The next two sections, Section \ref{trn} and Section \ref{NonEik}, are about the calculations of the some sub-leading eikonal corrections to the
BK equation where we derive these corrections for demonstrating the method of our calculations. The last Section summarizes our paper  where we also discuss the possible applications of the framework. 

\section{S-matrix element of quark's propagation in an external gluon field}
\label{SS}

 In order to understand an appearance of the Wilson line operators in the high energy scattering amplitudes let us consider a propagation
of quark in the external field, we do not consider here the same problem with gluons included. The task, of course, is well known, see \cite{Barba,ShurVain,Meggio,Laenen} references, 
additional examples can be found in \cite{Bal,Chiril,Zubkov} as well.
Introducing $J_{i}$ and $J_{f}$ as sources of quark's creation and absorption, we define the S-matrix element for the general case of quark's propagation in an external field:
\beq\label{SSec1}
S_{f i}\,= \,-\,\int\,d^{4}x_{i}\,d^{4}x_{f}\,J_{f}(x_{f})\,S(x_{f},\,x_{i})\,J_{i}(x_{i})\,,
\eeq
with the processes of creation and absorption of the quark in $y$ and $x$ correspondingly and $S(x, y)$ as quark's propagator in the external field.
For example, for the quark asymptotically free at $x,y\rightarrow\,\pm\,\infty$  we have:
\beq\label{SSec2}
J_{f}(x_{f})\,=\,\overl{u}(p)\,e^{\imath\,p\,x_{f}}\,\Le\,\imath\hat{\D}\,-\,m\,\Ra_{x_{f}}\,,\,\,\,
J_{i}(x_{i})\,=\Le\,\imath\cev{\hat{\D}}\,+\,m\,\Ra_{x_{i}}\,e^{-\imath\,q\,x_{i}}\,u(q)\,\,.
\eeq
Integration by parts provides in turn:
\beqar\label{SSec3}
S_{f i}\,&= &\,\overl{u}(p)\,\Le\hat{p}\,-\,m\,\Ra\,\int\,d^{4}x_{f}\,d^{4}x_{i}\,e^{\imath\,p\,x_{f}}\,S(x_{f},\,x_{i})\,e^{-\imath\,q\,x_{i}}\,
\Le\hat{q}\,-\,m\,\Ra\,u(q)\,=\,
\nonumber \\
&=&\,
\overl{u}(p)\,\Le\hat{p}\,-\,m\,\Ra\,S(p,\,q)\,\Le\hat{q}\,-\,m\,\Ra\,u(q)\,
\eeqar
with the limits $p^2,q^2\,\rightarrow\,m^2$ assumed to be taken at the end. Further we consider the processes with the quark propagation in an external field, 
the form of the propagator is complicated in comparison to the propagators of scalar or vector particles. Therefore, in order to make the calculations easier,
we will follow to \cite{Feynman} and
introduce the two component spinors: 
\beq\label{SSec6}
u(q)\,=\,\frac{1}{m}\,(\hat{q}\,+\,m)\,u(q)_{L}\,=\,\sqrt{m}\,\left (\begin{array}{c} 
1\\
\frac{q^{0}\,+\,\sigma^{i}\,q_{i}}{m}
\end{array}\right)\,\Psi_{L}(q)\,
\eeq
where correspondingly
\beq\label{SSec5}
u(q)_{L}\,=\,\frac{1}{2}(1\,-\,\gamma^{5})\,u(q)\,=\,\frac{\sqrt{m}}{2}(1\,-\,\gamma^{5})\,
\left (\begin{array}{c} 
\Psi_{L}(q)\\
\Psi_{R} (q)
\end{array}\right)\,=\,\sqrt{m}\,
\left (\begin{array}{c} 
\Psi_{L}(q)\\
0
\end{array}\right)\,
\eeq
with the chiral basis for the gamma matrices used:
\beq\label{SSec4}
\gamma^{\mu}\,=\,\left (
\begin{array}{cc} 
0 & \sigma ^{\mu}\\
\bar{\sigma}^{\mu} & 0
\end{array}\right)\,,\,\,\,\,\,
\sigma ^{\mu}\,=\,(1,\sigma^{i})\,,\,\,\,\,\,\bar{\sigma}^{\mu}\,=\,(1,-\sigma^{i})\,,\,\,\,\,\,
\gamma^{5}\,=\,\left (
\begin{array}{cc} 
-I & 0\\
0 & I
\end{array}\right)\,.
\eeq
Similarly we can consider a right-handed spinor:
\beq\label{SSec7}
u(q)_{R}\,=\,\frac{1}{2}(1\,+\,\gamma^{5})\,u(q)\,=\,\frac{\sqrt{m}}{2}(1\,+\,\gamma^{5})\,
\left (\begin{array}{c} 
\Psi_{L}(q)\\
\Psi_{R} (q)
\end{array}\right)\,=\,\sqrt{m}\,
\left (\begin{array}{c} 
0\\
\Psi_{R}(q)
\end{array}\right)\,
\eeq
with
\beq\label{SSec8}
u(q)\,=\,\frac{1}{m}\,(\hat{q}\,+\,m)\,u(q)_{R}\,=\,\sqrt{m}\,\left (\begin{array}{c} 
\frac{q^{0}\,-\,\sigma^{i}\,q_{i}}{m} \\
1
\end{array}\right)\,\Psi_{R}(q)\,.
\eeq
In any case, writing the \eq{SSec3} quarks propagator in the two component form
\beq\label{SSec9}
S(p,\, q)\,=\,\left (
\begin{array}{cc} 
G_{L}(p,q) & G_{1}(p,q)\\
G_{2}(p,q) & G_{R}(p,q)
\end{array}\right)\,,
\eeq
we will stay with only $G_{L}$ or $G_{R}$ component in the reduced \eq{SSec3} expression for the case of the propagation of a quark with 
chirality preservation\footnote{The change of chirality is provided by matrix element of the  $\overl{\Psi}_{R,L}\,\cdots\,\Psi_{L,R}$
with more complicated propagator, we do not consider this case in the paper.}.
So, let's rewrite the spinor in terms of $\Psi_{R}$ and defining
\beq\label{SSec10}
\overl{u}(p)\,=\,\sqrt{m}\,\overl{\Psi}_{R}(p)\,(\frac{p^{0}\,+\,\sigma^{i}\,p_{i}}{m}\,\,\,\,\,1)
\eeq
we write for the T-matrix element which corresponds to the \eq{SSec3} S-matrix element:
\beq\label{SSec11}
T_{f i}\,= \,\frac{1}{m}\,\overl{\Psi}_{R}(p)\,(p^2\,-\,m^2)\,\Le G_{R}(p,q)\,-\,G_{R\,0}(p,q)\,\Ra\,(q^2\,-m^2)\,\Psi_{R}(q)\,.
\eeq
here $G_{R\,0}$ is a free propagator at the absence of the external field.
In the case, more interesting for us, when the quark is created at some $y$ and asymptotically free at x, we will obtain instead \eq{SSec11}:
\beq\label{SSec111}
T_{f i}\,= \,-\,\frac{1}{m}\,\overl{\Psi}_{R}(p)\,(p^2\,-\,m^2)\,\int\,d^ 4 q\,\Le G_{R}(p,q)\,-\,G_{R\,0}(p,q)\,\Ra\,\tilde{J}_{i}(q)\,
\eeq
with properly redefined $J(q)$ from initial \eq{SSec1}.

 Therefore, we need to find the expression for $G_{R}(x,y)$  component of $S(x,\, y)$, we have for the four component propagator:
\beq\label{SSec12}
\Le \imath\,\hat{D}\,+\,\hat{\tv}\,-\,m \Ra\,\left (
\begin{array}{cc} 
G_{L} & G_{1}\\
G_{2} & G_{R}
\end{array}\right)\,=\,I_{4}\,\delta^{4}_{x y}\,,
\eeq
with $\tv$ as an external gluon field and a coupling constant included in it's definition. Correspondingly we obtain:
\beqar\label{SSec13}
&\,&
-m\,G_{1}\,+\,\imath\,\left[\,
\sigma^{0}\,\Le \D_{0} - \imath\,\tv_{0}\Ra\,-\,\sigma^{i}\,\Le \D_{i} - \imath\,\tv_{i}\Ra\,\right]\,G_{R}\,=\,0
\nonumber \\
&\,&
-m\,G_{R}\,+\,\imath\,\left[\,
\sigma^{0}\,\Le \D_{0} - \imath\,\tv_{0}\Ra\,+\,\sigma^{i}\,\Le \D_{i} - \imath\,\tv_{i}\Ra\,\right]\,G_{1}\,=\,I_{2}\,\delta^{2}\,.
\eeqar
Resolving this system we have for the $G_{R}$ propagator:
\beq\label{SSec1301}
\Le\,\left[\,
\sigma^{0}\,\Le \D_{0} - \imath\,\tv_{0}\Ra\,+\,\sigma^{i}\,\Le \D_{i} - \imath\,\tv_{i}\Ra\,\right]\,
\left[\,
\sigma^{0}\,\Le \D_{0} - \imath\,\tv_{0}\Ra\,-\,\sigma^{i}\,\Le \D_{i} - \imath\,\tv_{i}\Ra\,\right]\,
+\,m^2\Ra\,G_{R}\,=\,-\,\delta^{2}\,,
\eeq
where we made $G_{R}/m\,\rightarrow\,G_{R}$ substitution for the sake of simplicity, further we also will use the anti-hermitian representation of gluon field 
$\tv\,\rightarrow\,\imath\,\tv$.
Therefore we rewrite \eq{SSec13} as
\beq\label{SSec14} 
\Le
(\sigma^{0})^2\,\Le \D_{0} - \tv_{0}\Ra^2\,-\,\sigma^{0}\,\sigma^{i}\,F_{0 i}\,-\,\sigma^{i}\,\sigma^{j}\,
\Le \D_{i} - \tv_{i} \Ra\,\Le \D_{j} - \tv_{j} \Ra\,
+\,m^2\,\Ra\,G_{R}\,=\,-\,\delta^{2}\,
\eeq
or
\beq\label{SSec15} 
\Le
\Le \D_{\mu} - \tv_{\mu}\Ra^2\,-\,\sigma^{0}\,\sigma^{i}\,F_{0 i}\,-\,\frac{1}{4}\,[\sigma^{i}\,\sigma^{j}]\,F_{i j}\,
+\,m^2\,\Ra\,G_{R}\,=\,-\,\delta^{2}\,.
\eeq
Finally, defining
\beq\label{SSec16} 
\sigma^{\mu \nu}\,=\,\frac{1}{2}\,\Le \sigma^{\mu}\,\bar{\sigma}^{\nu}\,-\,\sigma^{\nu}\,\bar{\sigma}^{\mu}\Ra\,,
\eeq
see \eq{SSec4} definition, we obtain for the propagator:
\beq\label{SSec17} 
\Le
\Le \D_{\mu} - \tv_{\mu}\Ra^2\,+\,\frac{1}{2}\,\sigma^{\mu \nu}\,F_{\mu \nu}\,
+\,m^2\,\Ra\,G_{R}\,=\,-\,\delta^{2}\,.
\eeq
Our next step, consequently, is to write the propagator in the path integral form and isolated from it the requested part which reproduces the Wilson line currents in the effective action
of Lipatov. This task is important because whereas the presence of the simple ordered exponentials in the Lipatov's effective action is explained in the \cite{OurZub} paper, we face a problem
to account the sub-leading and non-eikonal correction there. Now, because the effective currents in the action can be interpreted as a contribution from the truncated propagators, 
using the precise form of the truncated propagator we consequently can account the sub-leading and non-eikonal corrections to the processes described by initial Lipatov's action.

 The different presentation of the quark's propagator in an external field in the form of the path integral can be found in \cite{Barba,Meggio,Laenen,Chiril}, see also \cite{Rosen,Venez} for examples, 
partially we will use these results. We begin from the standard representation of the \eq{SSec17} propagator: 
\beq\label{SSec18} 
G_{R}(x_{f},x_{i})\,=\,-\,\frac{\imath}{2}\,\int_{T_{0}}^{\infty}\,d T\,\int^{x_{f}=z(T)}_{x_{i}=z(T_{0})} 
\,\mathcal{D} x\,\,PExp\left[ \imath\int_{T_{0}}^{T}\,dt\, \Le
\frac{\dot{x}^2}{2}\,-\,\frac{m^2}{2}\,+\,\dot{x}_{\mu}\,\tv^{\,\mu}\,+\,\frac{1}{4}\,\sigma^{\mu \nu}\,F_{\mu \nu}\Ra\,\right]\,.
\eeq
Assuming parametrization
\beq\label{SSec19} 
\,x\,=\,x_{straight}\,+\,\xi(t)\,=\,
x_{i}\,+\,(x_{f}-x_{i})\,\frac{t-T_{0}}{T-T_{0}}+\,\xi(t)\,,\,\,\,\xi(T)\,=\,\xi(T_{0})\,=\,0\,
\eeq
or the equivalent
\beq\label{SSec191} 
\,x\,=\,x_{straight}\,+\,\xi(t)\,=\,C\,+\,p\,t\,+\,\xi\,,\,\,\,\dot{C}\,=\,0\,,
\eeq
and extracting the bare propagator from the expression, i.e. leaving in the expression only part of the propagator which depends on the
external field, we obtain:
\beqar\label{SSec20} 
&\,&\hat{G}_{R}(x_{f},x_{i})\,=\,
-\,\frac{\imath}{2}\,\int_{T_{0}}^{\infty}\,d T\,e^{\frac{\imath}{2}\Le p^2 -m^2\Ra\,\Le T-T_{0}\Ra\,}
\int \mathcal{D} \xi\,
e^{\frac{\imath}{2}\,\int_{T_{0}}^{T}\,dt\,\dot{\xi}^2\,}\,
\nonumber \\
&\,&
\Le\,P Exp\left[ \imath\int_{T_{0}}^{T}\,dt\, \Le
\,\Le p^{\mu}+\dot{\xi}^{\mu}\Ra\,\tv_{\,\mu}(x)\,+\,\frac{1}{4}\,\sigma^{\mu \nu}\,F_{\mu \nu}(x)\Ra\,\right]\,-\,1\Ra \,=\,
\nonumber \\
&=&
-\,\frac{\imath}{2}\,\int_{T_{0}}^{\infty}\,d T\,e^{\frac{\imath}{2}\Le p^2 -m^2\Ra\,\Le T-T_{0}\Ra\,}
\int \mathcal{D} \xi\,
e^{\frac{\imath}{2}\,\int_{T_{0}}^{T}\,dt\,\dot{\xi}^2\,}\,
\nonumber \\
&\,&
\Le\,P Exp\left[\imath\int_{x(T_{0})=x_{i}}^{x(T)=x_{f}}\,dx^{\mu}\,\tv_{\,\mu}(x)\,+\, \imath\int_{T_{0}}^{T}\,dt\, \Le
\dot{\xi}^{\mu}\,\tv_{\,\mu}(x)\,+\,\frac{1}{4}\,\sigma^{\mu \nu}\,F_{\mu \nu}(x)\Ra\,\right]\,-\,1\Ra \,,
\eeqar
here the substitution $p\,=\,\dot{x}_{straight}$ was made. This substitution can be understood as a straight line approximation for the description of the high energy quark's propagation in an external field. Namely, the $dx^{\mu}$ integral in the \eq{SSec20} exponential must be defined as an integral on the 
the straight line quark trajectory, see Section~\ref{NonEik} further.

 Considering the propagation of the quark free at $x$, see \eq{SSec111}, we need an expression 
with the $(p^2-m^2)^{-1}$ pole isolated, see \cite{Laenen} for additional examples. Proceeding with \eq{SSec20}, we restore the regularization of the \eq{SSec15} operator adding
$\imath\,\varepsilon$ term in the inverse propagator, and write for the factor which determines the answer for the \eq{SSec111}:
\beqar\label{SSec21} 
&\,&\hat{T}_{f i}\,=\,-\,(p^2-m^2)\,(p^2-m^2)^{-1}\,
\int_{T_{0}}^{\infty}\,d T\,\Le \frac{d}{dT}\,e^{\frac{\imath}{2}\Le p^2 -m^2\Ra\,\Le T-T_{0}\Ra\,}\Ra
\int \mathcal{D} \xi\,
e^{-\frac{1}{2}\,\varepsilon\,(T-T_{0})\,+\,\frac{\imath}{2}\,\int_{T_{0}}^{T}\,dt\,\dot{\xi}^2\,}\,
\nonumber \\
&\,&
\Le\,P Exp\left[ \imath\int_{T_{0}}^{T}\,dt\, \Le
\,\Le p^{\mu}+\dot{\xi}^{\mu}\Ra\,\tv_{\,\mu}\,+\,\frac{1}{4}\,\sigma^{\mu \nu}\,F_{\mu \nu}\Ra\,\right]\,-\,1\Ra \,=\,
\nonumber \\
&=&\,
-\,
\int_{T_{0}}^{\infty}\,d T\,\Le \frac{d}{dT}e^{\frac{\imath}{2}\Le p^2 -m^2\Ra\,\Le T-T_{0}\Ra\,}\Ra\,e^{-\frac{1}{2}\,\varepsilon\,(T-T_{0})}\,f(T_{0},\,T)\,.
\eeqar
After the integration by parts we obtain:
\beq\label{SSec22} 
\hat{T}_{f i}\,=\,-\,\Le -\,f(T_{0},\,T_{0})\,-\,\int_{T_{0}}^{\infty}\,d T\,
e^{\frac{\imath}{2}\Le p^2 -m^2\Ra\,\Le T-T_{0}\Ra\,}\,\frac{df(T,\,T_{0})}{dT}\,\Ra\,.
\eeq
Now we can take $p^2\,\rightarrow\,m^2$ limit in the exponential in the integral obtaining finally
\beq\label{SSec2201} 
\hat{T}_{f i}\,=\,f(\infty,\,T_{0})\,
\eeq
with
\beqar\label{SSec23}
&\,& f(\infty,\,T_{0})\,= \,
\int \mathcal{D} \xi\,e^{\frac{\imath}{2}\,\int_{T_{0}}^{\infty}\,dt\,\dot{\xi}^2}\,
\Le\,P Exp\left[ \imath\,\int_{T_{0}}^{\infty}\,dt\, \Le\,
\,\Le p^{\mu}+\dot{\xi}^{\mu}\Ra\,\tv_{\,\mu}\,+\,\frac{1}{4}\,\sigma^{\mu \nu}\,F_{\mu \nu}\Ra\,\right]\,-\,1\Ra \,=\,
\nonumber \\
&=&
\int \mathcal{D} \xi
e^{\frac{\imath}{2}\int_{T_{0}}^{T} dt \dot{\xi}^2}
\Le P Exp\left[\imath\int_{x_{i}}^{x_{f}} dx^{\mu} \tv_{\,\mu}(x)+\imath\int_{T_{0}}^{\infty} dt \Le
\dot{\xi}^{\mu} \tv_{\,\mu}(x) +\frac{1}{4}\sigma^{\mu \nu} F_{\mu \nu}(x)\Ra \right] - 1\Ra \,,
\eeqar
see also \cite{Laenen}. We consider this "phase" operator is a main object of our interests, after the proper normalization  it coincides to LO precision with the simplest Wilson line 
operator. The \eq{SSec23} expression can be directly used in the effective action of Lipatov, see further. From the expression it is clear, that the 
form of the same object for the gluon propagation in the external field
will coincide with the \eq{SSec23} only to the LO precision providing different sub-leading and non-eikonal corrections.

  A different answer we will obtain if we will interesting in the T-matrix element for the quark free at $x_{i}$ and $x_{f}$, in this case we need to isolate two poles
$(p^2-m^2)^{-1}$ and $(q^2-m^2)^{-1}$, see \eq{SSec11} expression. We again begin from \eq{SSec20} and expanding P-exponential rewrite it as following:
\beqar\label{SSec24} 
&\,&
\hat{G}_{R}(x_{f},x_{i})\,=\,
\frac{1}{2}\int_{T_{0}}^{\infty}\,d T\,e^{\frac{\imath}{2}\Le p^2 -m^2\Ra \Le T-T_{0}\Ra}\,
 \int \mathcal{D} \xi\,
e^{\frac{\imath}{2}\,\int_{T_{0}}^{T}\,d\eta\,\dot{\xi}^2\,}\,\int_{T_{0}}^{T}\,dt\,\Le \D_{t}\,P e^{\imath\int_{T_{0}}^{t}\,\phi(\xi (\eta))\,d\eta}\,\Ra\,=\,
\nonumber \\
&=&
\frac{1}{2}\int_{T_{0}}^{\infty}\,d T\,e^{\frac{\imath}{2}\Le p^2 -m^2\Ra \Le T-T_{0}\Ra}\,
 \int \mathcal{D} \xi\,
e^{\frac{\imath}{2}\,\int_{T_{0}}^{T}\,d\eta\,\dot{\xi}^2\,}\,\int_{T_{0}}^{T}\,dt\,\phi(\xi (t))
\, P e^{\imath\int_{T_{0}}^{t}\,d\eta\, \phi(\xi (\eta))}\,
\eeqar
with correspondingly defined $\phi$ function. At the next step we perform variables change in \eq{SSec3} and \eq{SSec24}: 
\beqar\label{SSec25}
\dot{\xi}\,&=&\,\dot{\tilde{\xi}}\,+\,(q\,-\,p)\,\theta(t\,-\,\eta)\,, \\
x_{i}\,& = &\,\tilde{x}_{i}\,+\,(q\,-\,p)\,\Le t\,-\,T_{0}\,\Ra\,.
\eeqar
Therefore we will obtain for \eq{SSec24} expression:
\beq\label{SSec26} 
\hat{G}_{R}(x_{f},x_{i})\,=\,
\frac{1}{2}\int_{T_{0}}^{\infty}\,d T\,\int_{T_{0}}^{T}\,dt \,e^{\frac{\imath}{2}\Le p^2 -m^2\Ra \Le T+t-2 T_{0}\Ra
-\frac{\imath}{2}\Le q^2 -m^2\Ra \Le t- T_{0}\Ra}
\int \mathcal{D} \xi
e^{\frac{\imath}{2}\,\int_{T_{0}}^{t}\,d\eta\,\dot{\tilde{\xi}}^2\,}\,\phi
\, P e^{\imath\int_{T_{0}}^{t}\,d\eta\, \phi}\,,
\eeq
with properly redefined argument of $\phi$ function. Redefining the variables in the integral we correspondingly can write:
\beq\label{SSec27} 
\hat{G}_{R}(x_{f},x_{i})\,=\,
\frac{1}{2}\int_{0}^{\infty}\,d T\,\int_{0}^{T}\,dt \,e^{\frac{\imath}{2}\Le p^2 -m^2\Ra \Le T+t\Ra
-\frac{\imath}{2}\Le q^2 -m^2\Ra t}
\int \mathcal{D} \xi
e^{\frac{\imath}{2}\,\int_{0}^{t}\,d\eta\,\dot{\tilde{\xi}}^2\,}\,\phi
\,e^{\imath\int_{0}^{t }\,d\eta\, \phi}\,
\eeq
and proceeding similarly to done in \eq{SSec21} we again will obtain an expression for the T-matrix element, see \cite{Barba,Rosen}.

\section{Perturbative calculation of Wilson lines correlators}
\label{WL}

 Our next step is a consideration of the variant of the  Lipatov's effective action written in the form of interacting Wilson lines: 
\beq\label{WL2}
Z\,= \,
\frac{1}{Z^\prime} \int \mathcal{D}\tv \,\, {\rm exp}\,\Big(\imath\, S^{0}[\tv]\,-\,
\frac{\imath}{2\,C(R)}\int\, d^4 x \,\T_{+}\,\partial_{\bot}^{2}\,\T_{-}\,\Big)\
\eeq
here $S^{0}$ as the QCD gluon's action on which the averaging of the operators is performed, $C(R)$ is an eigenvalue of the Casimir operator for the corresponding gluons representation and
\beq\label{Add3}
\T_{\pm}(\textsl{v}_{\pm})\,=\,\frac{1}{g}\,\D_{\pm}\,O(\textsl{v}_{\pm})\,=\,
\textsl{v}_{\pm}\,O(\textsl{v}_{\pm})\,,
\eeq
\beq\label{Add4}
O(\tv_{\pm})\,=\,P\,e^{g\,\int_{-\infty}^{x^{\pm}}\,d x^{\pm}\,\tv_{\pm}(x^{\pm},\,\,x_{\bot})}\,,\,\,\,\,\tv_{\pm}\,=\,\imath\,T^{a}\,\tv_{\pm}^{a}\,.
\eeq
This version of the Lipatov's action was derived in \cite{OurZub,Our4}, it's form is different from the initial Lipatov's action but, as was demonstrated in \cite{OurZub}, it is fully equivalent to it. 
The use of this variant of the action is dictated by the presence of the Wilson lines operators in this variant of the action, 
they are main degrees of freedom in the BK-JIMWLK approaches as well.
Therefore, it allows relatively simple comparison between the approaches, in all of them we will operate with the same
type of the degrees of freedom.

  Now, introducing auxiliary currents and applying the light-cone gauge, we obtain:
\beq\label{WL12}
Z[J] \,= \,
\frac{1}{Z^\prime} \int D\tv \, {\rm exp}\,\Big(\imath\, S^{0}[\tv]\,-\,
\frac{\imath}{2\,C(R)}\int\, d^4 x \,\Le \T_{+}- J_{+}(x^+, x_{\bot})\,\partial_{\bot}^{-2}\Ra \, J_{-}(x^-, x_{\bot})\,.
\Big)\
\eeq
In the following we consider the Regge like kinematics of the interactions, i.e. shock wave approximation of the scattering. 
Therefore, we define the currents as following:
\beq\label{WL122}
\int dx^{-}\,J_{-}(x^-, x_{\bot})\,\rightarrow\,\int dx^{-}\,\delta(x^{-})J_{-}\,(x_{\bot})\,.
\eeq
Taking two derivatives of $\log Z[J]$ with respect to the currents we have:
\beqar\label{WL21}
&\,&\,-\,C(R)^2\,(\,2\,\g)^{2}\,\Le\,\frac{\delta^{2}}{\delta J_{1\,-}^{a_1}\,\delta J_{2\,-}^{a_2}}\,\log Z[J]\,\Ra_{\,J\,=\,0}\,=\, 
\int dx^{-}\,\delta(x^{-})\,\int dy^{-}\,\delta(y^{-})
\,\nonumber \\
&\,&\,<\,T^{a_1}\,\Le O_{1}(\tv_{+})_{x^{+}\,=\,\infty}\,-\,O_{1}(\tv_{+})_{x^{+}\,=\,-\infty}\, \Ra\,\otimes\,T^{a_2}
\Le O_{2}(\tv_{+})_{x^{+}\,=\,\infty}\,-\,O_{2}(\tv_{+})_{x^{+}\,=\,-\infty}\, \Ra\,>\,=\, \nonumber \\
&=&\,
\int dx^{-}\,\delta(x^{-})\,\int dy^{-}\,\delta(y^{-})\,
<\,\Le T^{a_1}\,V_{\,1}(\tv_{+})\Ra\,\otimes\,\Le T^{a_2}\,V_{\,2}(\tv_{+})\Ra\,>\,,
\eeqar
further, in sake of simplicity, we will not write the integrals on $x^{-}, y^{-}$ variables precisely and will account the 
integrals only in the calculations of the final contributions to BK equation.
The \eq{WL21} correlator is defined through the usual Wilson lines:
\beq\label{WL2101}
\hat{O}(\tv_{\pm})\,=\,V(\tv_{\pm})\,=\,P\,e^{g\,\int_{-\infty}^{\infty}\, d x^{\pm}\,\tv_{\pm}(x^{\pm},\,x^{\mp},\,x_{\bot})}\,-\,1\,,\,\,\,\,\tv_{\pm}\,=\,\imath\,T^{a}\,\tv_{\pm}^{a}\,.
\eeq
We note, that the form of the effective action, \eq{WL2}, allows to rewrite the $\T_{\pm}(\textsl{v}_{\pm})$ operators in terms of
\eq{SSec23} phase factor, it can be included directly
in the \eq{WL2} action. 
In this case, the action allows systematically account
sub-leading and sub-eikonal correction to the \eq{WL21} correlator.
We have to mention also, that the \cite{Bal} approach operates with the effective current in the action given by the quark-antiquark (quark) loop, i.e. by the physical currents which include full quark's propagators, consequently in the expression for the amplitude the corresponding impact-factor is arising. 
In contrast to that, the effective currents in the Lipatov's effective action are untended to reconstruct
the correct BFKL correlators of reggeized gluons with proper rapidity dependence, see \cite{Our1,Our2}. Therefore, whereas in the Balitsky's approach we need to know the form of the whole quark's propagator, 
in the form \eq{SSec18} for example, in the Lipatov's approach we need a "phase" operator in the form of \eq{SSec23}, with the properly defined normalization of the path integral on $\xi$ 
variable. This issue we also discuss in the \ref{NonEik} Section.

  Now, taking the gluon field as the classical solution (Reggeon field) of the equations of motion of the effective action of Lipatov, see \cite{Our1,Our2}, plus the quantum fluctuation around
\beq\label{N1}
\tv_{+}\,=\,\B_{+}(x^{+},x_{\bot})\,+\,\ep_{+}\,
\eeq
we can expand the $V$ operator in respect to these fluctuations:
\beqar\label{N2}
V(\tv_{+})\,& = &\,V(\B_{+}(x^{+},x_{\bot}))\,+\,g\,\int^{\infty}_{-\infty}\,dx^+\,\Le O^{T}(\B_{+}(x^{+},x_{\bot}))\,
\ep_{+}(x)\,O(\B_{+}(x^{+},x_{\bot}))\,\Ra\,+\,\nonumber\\
&+&\,
\frac{g^2}{2}\,\int^{\infty}_{-\infty}\,dx^+\,O^{T}(\B_{+}(x^{+},x_{\bot}))\,\ep_{+}(x)\,\int d^4 p\,
G^+\,(x,p)\,\ep_{+}(p)\,O(\B_{+}(p^{+},p_{\bot}))\,+\,\nonumber \\
&+&\,
\frac{g^2}{2}\,\int^{\infty}_{-\infty}\,dx^+\,\int d^4 p\,O^{T}\,(\B_{+}(p^{+},p_{\bot}))\,\ep_{+}(p)\,
 G^+\,(p,x)\,\ep_{+}(x)\,O(\B_{+}(x^{+},x_{\bot}))+\cdots\,.
\eeqar
Here we defined
\beqar\label{N3}
&\,& G_{x y}^{+}\,= \,G_{x y}^{+\,0}\,+\,g\,G_{x z}^{+\,0}\, v_{+ z}\,G_{z y}^{+}\,,\,\,\,D_{+\,x y}\,G_{y z}^{+}\,=\,\delta^{4}_{x z} \,\\
&\,&G_{x y}^{+\,0}\,= \,\theta(x^{+}\,-\,y^{+})\,\delta^{3}_{x y}\,,\,\,\,\D_{+\,x}\,G_{x y}^{+\,0}\,=\,\delta^{4}_{x y}\, 
\label{N301}\\
&\,& O^{T}(\tv_{\pm})\,= \,P\,e^{g\,\int^{\infty}_{x^{\pm}}\,d x^{\pm}\,\tv_{\pm}(x^{\pm},\,\,x_{\bot})}\,
\label{N302},
\eeqar
with $D$ as covariant derivative operator.
Finally we can define the correlator of interest writing precisely the LO terms only in the expression:
\beqar\label{N22}
&\,& < T^{a}\,V(x)\otimes T^{b} V(y) >= \Le  T^{a} V(x_{\bot})\Ra_{i i}\,\Le T^{b} V(y_{\bot})\Ra_{j j} \,+\, \nonumber \\
&+&
g\,\Le T^{a} V(x_{\bot}) \Ra_{i i}\,
\int^{\infty}_{-\infty}\,dy^+\,\Le T^{b}\,O^{T}_{y}\,(\imath\,T^{c})
\,O_{y}\Ra_{j j}\,< \ep_{+}^{c}(y) >\,+\, \nonumber \\
&+&
g\,
\int^{\infty}_{-\infty}\,dx^+\,\Le T^{a}\,O^{T}_{x}\,(\imath\,T^{c})
\,O_{x}\Ra_{i i}\,< \ep_{+}^{c}(x) >\,\Le T^{b} V(y_{\bot})\Ra_{j j}\,+\,\nonumber \\
&+&
\frac{g^2}{2}
\int^{\infty}_{-\infty}\,dx^+\,\int d^4 p\,\Le T^{a} O^{T}_{x}\,(\imath\,T^{c})\,
G^{+}(x,p)\,(\imath\,T^{d})\,O_{p}\Ra_{ii} \,<\ep_{+}^{c}(x) \ep_{+}^{d}(p)>
\Le T^{b} V(y_{\bot}) \Ra_{jj} +  \nonumber \\
&+&
\frac{g^2}{2}
\int^{\infty}_{-\infty}\,dx^+\,\int d^4 p\,\Le T^{a} O^{T}_{p}\,(\imath\,T^{c})\,
G^{+}(p,x)\,(\imath\,T^{d})\,O_{x}\Ra_{i i } \,<\ep_{+}^{c}(p) \ep_{+}^{d}(x)>
\Le T^{b} V(y_{\bot}) \Ra_{jj}+  \nonumber \\
&+&
\frac{g^2}{2}\Le T^{a} V(x_{\bot})\Ra_{ii} \, 
\int^{\infty}_{-\infty}\,dy^+\,\int d^4 p\,\Le T^{b} O^{T}_{p}\,(\imath\,T^{c})\,
 G^+\,(p,y)\,(\imath\,T^{d})\,O_{y}\Ra_{j j }\,<\ep_{+}^{c}(p) \ep_{+}^{d}(y)> +\nonumber \\
&+&
\frac{g^2}{2} \Le T^{a} V(x_{\bot}) \Ra_{ii }\,
\int^{\infty}_{-\infty}\,dy^+\,\int d^4 p\,\Le T^{b} O^{T}_{y}\,(\imath\,T^{c})\,
 G^+\,(y,p)\,(\imath\,T^{d})\,O_{p}\Ra_{jj}\,<\ep_{+}^{c}(y) \ep_{+}^{d}(p)> +\nonumber \\
&+&
g^2 
\int^{\infty}_{-\infty}\,dx^+\,\Le T^{a} O^{T}_{x}\,
(\imath\,T^{c}) \,O_{x}\,\Ra_{i i}\,
\int^{\infty}_{-\infty}\,dy^+\,\Le T^{b} O^{T}_{y}\,(\imath\,T^{d})
\,O_{y}\Ra_{j j}\,< \ep_{+}^{c}(x) \ep_{+}^{d}(y) >\,+\,\cdots\,.
\eeqar
The same we can write in the matrix form:
\beqar\label{N221}
&\,& < \,V(x)\otimes  V(y) >=  V_{i k}(x_{\bot})\, V_{l j}(y_{\bot}) \,+\, \nonumber \\
&+&
g\, V_{i k}(x_{\bot}) \,
\int^{\infty}_{-\infty}\,dy^+\,\Le \,O^{T}_{y}\,(\imath\,T^{c})
\,O_{y}\Ra_{l j}\,< \ep_{+}^{c}(y) >\,+\, \nonumber \\
&+&
g\,
\int^{\infty}_{-\infty}\,dx^+\,\Le \,O^{T}_{x}\,(\imath\,T^{c})
\,O_{x}\Ra_{i k}\,< \ep_{+}^{c}(x) >\, V_{l j}(y_{\bot})\,+\,\nonumber \\
&+&
\frac{g^2}{2}
\int^{\infty}_{-\infty}\,dx^+\,\int d^4 p\,\Le O^{T}_{x}\,(\imath\,T^{c})\,
G^{+}(x,p)\,(\imath\,T^{d})\,O_{p}\Ra_{ik} \,<\ep_{+}^{c}(x) \ep_{+}^{d}(p)>
\, V_{lj}(y_{\bot})\, +  \nonumber \\
&+&
\frac{g^2}{2}
\int^{\infty}_{-\infty}\,dx^+\,\int d^4 p\,\Le O^{T}_{p}\,(\imath\,T^{c})\,
G^{+}(p,x)\,(\imath\,T^{d})\,O_{x}\Ra_{i k } \,<\ep_{+}^{c}(p) \ep_{+}^{d}(x)>\,
V_{l j}(y_{\bot}) \,+  \nonumber \\
&+&
\frac{g^2}{2}\,V_{ik}(x_{\bot}) \, 
\int^{\infty}_{-\infty}\,dy^+\,\int d^4 p\,\Le  O^{T}_{p}\,(\imath\,T^{c})\,
 G^+\,(p,y)\,(\imath\,T^{d})\,O_{y}\Ra_{l j }\,<\ep_{+}^{c}(p) \ep_{+}^{d}(y)> +\nonumber \\
&+&
\frac{g^2}{2} \, V_{ik }(x_{\bot})\,
\int^{\infty}_{-\infty}\,dy^+\,\int d^4 p\,\Le  O^{T}_{y}\,(\imath\,T^{c})\,
 G^+\,(y,p)\,(\imath\,T^{d})\,O_{p}\Ra_{lj}\,<\ep_{+}^{c}(y) \ep_{+}^{d}(p)> +\nonumber \\
&+&
g^2 
\int^{\infty}_{-\infty}\,dx^+\,\Le O^{T}_{x}\,
(\imath\,T^{c}) \,O_{x}\,\Ra_{i k}\,
\int^{\infty}_{-\infty}\,dy^+\,\Le  O^{T}_{y}\,(\imath\,T^{d})
\,O_{y}\Ra_{l j}\,< \ep_{+}^{c}(x) \ep_{+}^{d}(y) >\,+\,\cdots\,.
\eeqar
This system of equations is similar (almost) to the  Balitsky, \cite{Bal}, hierarchy of the correlators. There is a complete system of equations for the 
Wilson line correlators written in terms of the correlators itself and correlators of gluon and 
quark\footnote{We can include the quarks in the action obtaining the additional contributions in the system
of the equations.} fields. Nevertheless, we do not assume nor any special forms of gluon's correlators in \eq{N221} neither any kind of large $N_c$ approximations, in general the system can not be 
be reduced directly to the standard BK equation.
For example, the 
$ \Le O^{T}_{x}\,T^{c} \,O_{x}\,\Ra_{i k}$ operator can not be written through \eq{WL21} Wilson lines which are present in the l.h.s. of the equation in the general case.
Therefore, further approximations must be done in the framework if we want to reproduce the standard Balitsky equation. 
But, before to proceed with the shock wave approximation\footnote{Further we will discuss the reduction of the \eq{N221} to the form of the Balitsky equation, we do not consider the large $N_c$
approximation and do not discuss the \cite{Kovch} Kovchegov's form of the equation.}, in the next Section we discuss the form of the gluon's propagator which appears
in the r.h.s. of \eq{N221}.

\section{Propagator of gluon field}
\label{Ef}

 In this Section we remind the framework of  \cite{Our2,Our5} and calculate the gluon field propagator in the shock wave  which we will
use in the derivation of BK equation further.  
First of all, we consider an expansion of the gluon field around classical solution 
provided by Lipatov's effective action in the light-cone gauge:
\beqar\label{Ef1}
\tv_{+}^{\,a} & = & \tv_{+}^{\,a cl}+\ep_{+}^{\,a}=\B_{+}^{\,a}(x^{+},x_{\bot})+\ep_{+}^{\,a}\, \\
\tv_{i}^{\,a} & = & \tv_{i}^{\,a cl}+\ep_{i}^{\,a} =
tr\,[f^{\,a}\,O(x^{+})
f^{\,b}\,O^{T}(x^{+})]\,\rho_{\,b i}\Le x^{-} , x_{\bot}\Ra +\ep_{i}^{\,a}\,,\,\,\,\,
\rho_{i}^{b}\,=\,
\Le\,\D_{-}^{-1}\,\Le\D_{i}\,\B_{-}^{\,b}\Ra\,\Ra\,,
\eeqar
here $\B_{\pm}(x^{\pm},\,x_{\bot})$ are the Reggeon fields,
see \cite{Our1,Our2,Our3,Our4,Our5} papers.
Expanding the QCD Lagrangian with respect to the fluctuations, we obtain the following expression:
\beqar\label{Ef2}
S_{\ep^{2}}& = &-\frac{1}{2}\,\int\,d^{4} x\,\left.\Big( \ep_{i}^{a}\Le \delta_{ac}\Le \delta_{ij}\,\Box\, +\D_{i}\,\D_{j} \Ra - \right.\right. \nonumber \\
&-&
2 g f_{abc} \Le \delta_{ij} \,\tv_{k}^{b\,cl} \D_{k}  -
2\, \tv_{j}^{b\,cl} \D_{i}  + \tv_{i}^{b\,cl} \D_{j} - \delta_{ij} \,\tv_{+}^{b\,cl} \D_{-}  \Ra    -\nonumber\\
&-&\left.\,
\,g^{2}\,f_{a b c_{1}}\,f_{c_{1} b_{1} c}\,
\Le \delta_{i j}\,v_{k}^{b\,cl}\,\tv_{k}^{b_{1}\,cl}\,
-\,\tv_{i}^{b_1\,cl}\,\tv_{j}^{b\,cl}\,\Ra\,\Ra\,\ep_{j}^{c}\,+\,\nonumber\\
&+&\,
\,\ep_{+}^{a}\Le -2 \,\delta^{a c}\,\D_{-}\D_{i} -2 g f_{abc} \Le  \tv_{i}^{b\,cl} \D_{-} - \Le \D_{-} \tv_{i}^{b\,cl}\Ra \Ra \Ra\,\ep_{i}^{c}\,
+\,
\left.\,
\ep_{+}^{a}\, \delta_{a c}\, \D_{-}^{2} \,\ep_{+}^{c} \,\Ra
\,=\,\nonumber \\
&=&\,-\,\frac{1}{2}\,\ep_{\mu}^{a}\,\Le\,\Le M_{0} \Ra_{\mu\, \nu}^{ac}\,+\,\Le M_{1} \Ra_{\mu\, \nu}^{ac}\,+\,
\Le M_{2} \Ra_{\mu\, \nu}^{ac}\,\Ra\,\ep_{\nu}^{c}\,.
\eeqar
Here we defined $\Le M_{i} \Ra_{\mu\, \nu}^{ac}\,\propto\,g^{i}$ and note that
\beq\label{Ef3}
\Le M_{1}\Ra_{-\,i}\,=\,-\, g f_{abc} \Le  \tv_{i}^{b\,cl} \,\overrightarrow{\D_{-}} - \Le \D_{-} \tv_{i}^{b\,cl}\Ra \Ra\,,\,\,\,
\Le M_{1}\Ra_{i\,-}\,=\,-\, g f_{abc} \Le  \,\overleftarrow{\D_{-}}\,  \tv_{i}^{b\,cl}- \Le \D_{-} \tv_{i}^{b\,cl}\Ra \Ra\,.
\eeq
The bare  gluon propagator $\,G_{\,0\,\nu\,\mu}$ is defined as
\beq\label{Ef6}
\Le M_0\Ra_{\,\mu\,\nu}\,G_{\,0\,\nu\,\rho}\,=\,\delta_{\mu \rho}\,,
\eeq
and correspondingly the full gluon propagator we write as following:
\beq\label{Ef7}
G_{\mu \nu}^{ac}\,=\,\left[\,\Le M_{0} \Ra_{\mu \nu}^{ac}\,+\,\Le M_{1} \Ra_{\mu \nu}^{ac}\,+\,\Le M_{2} \Ra_{\mu \nu}^{ac}\,
\,\right]^{-1}\,.
\eeq
The \eq{Ef7} also can be written in the form of the following  perturbative series:
\beq\label{Eff8}
G_{\mu \nu}^{ac}(x,y)\,=\,G_{0\,\mu \nu}^{ac}(x,y)\,-\,\int\,d^4 z\,G_{0\,\mu \rho}^{ab}(x,z)\,
\Le \,\Le M_{1}(z)\Ra_{\rho \gamma}^{bd}\,+\,\Le M_{2}(z)\Ra_{\rho \gamma}^{bd}\Ra G_{\gamma \mu}^{dc}(z,y)\,.
\eeq
To the first order of the approximation we can take $v_{i}^{b\,cl}\,=\,0$. Therefore, the correlator (propagator) we need has the following form:
\beqar\label{Eff9}
G_{+ +}^{ac}(x,y)\,& = &\,G_{0\,+ +}^{ac}(x,y)\,-\,\int\,d^4 z\,G_{0\,+ i}^{ab}(x,z)\,
 \,\Le M_{1}(z)\Ra_{i j}^{bd} \, G_{j +}^{dc}(z,y)\, \\
G_{j +}^{ac}(x,y)\,& = &\,G_{0\,j +}^{ac}(x,y)\,-\,\int\,d^4 z\,G_{0\,j i}^{ab}(x,z)\,
 \,\Le M_{1}(z)\Ra_{i k}^{bd} \, G_{k +}^{dc}(z,y)\,\label{Eff91}
\eeqar
Here
\beq\label{Eff10}
\Le M_{1}(z)\Ra_{i j}^{bd}\,=\,- 2 g\,f_{b d w }\,\delta_{ij} \,v_{+}^{w\,cl} \D_{-}= -
2 \, g\,\delta_{ij} \,v_{+}^{\,cl} \D_{-}\,,\,\,\,\,v_{+}^{\,cl}\,=\,\imath\,T^a\,v_{+}^{a\,cl} ,
\eeq
\beq\label{Eff11}
G_{0\,+\,+}(x,y)\,=\,-\,\int\,\frac{d^4 p}{(2\pi)^{4}}\,\frac{e^{-\imath\,p\,(x\,-\,y)}}{p^{\,2}}\,\frac{2\,p_{\,+}}{p_{\,-}}\,,
\eeq
\beq\label{Eff12}
G_{\,0\,i\,+}\,=\,G_{\,0\,+\,i}\,=\,\int\,\frac{d^4 p}{(2\,\pi)^{4}}\,\frac{e^{-\,\imath\,p\,(x\,-\,y)}}{p^{\,2}}\,
\frac{p_{\,i}}{p_{-}}\,,
\eeq	
\beq\label{Eff13}
G_{0\,i\,j}(x,y)\,=\,-\,\int\,\frac{d^4 p}{(2\pi)^{4}}\,\frac{e^{-\imath\,p\,(x\,-\,y)}}{p^{\,2}}\,\delta_{\,i\,j}\,.
\eeq
Now we reproduce the Balitsky-Belitsky result, see \cite{BalBel}, performing the full re-summation of \eq{Eff9} expression using  the $v_{+}^{cl}\,=\,\B_{+}$ classical field. Firstly consider the first order contribution of the classical field to \eq{Eff9} propagator:
\beqar\label{CO1}
&\,& I_1 = \Le -2\,\imath\, g \Ra\,\int d^4 z\,\int \frac{d^4 p}{(2\pi)^4} \frac{e^{-\,\imath\,p\,(x\,-\,z)}}{p^{\,2}}\frac{p_{\,i}}{p_{-}}\,
\int \frac{d^4 k}{(2\pi)^4} \frac{e^{-\,\imath\,k\,(z\,-\,y)}}{k^{\,2}}\,k_{\,i}\,\tv^{\,cl}_{+}(z)\,=\, \\
&=&
\Le -2\,\imath\,g \Ra\,\int \frac{d^4 p}{(2\pi)^4} \frac{e^{-\,\imath\,p\,x\,}}{p^{\,2}}\frac{p_{\,i}}{p_{-}}\,
\int \frac{d^4 k}{(2\pi)^4} \frac{e^{\,\imath\,k\,y}}{k^{\,2}}\,k_{\,i}\,
\int d^4 z\,e^{\,\imath\,z\,(p\,-\,k)}\,\tv^{\,cl}_{+}(z)\,=\,\Le -2\imath g \Ra 2 \pi\,\nonumber \\
&\,&
\int \frac{d^4 p}{(2\pi)^4} \frac{e^{-\,\imath\,p\,x\,}}{p^{\,2}}\frac{p_{\,i}}{p_{-}}
\int \frac{d^4 k}{(2\pi)^4} \frac{e^{\,\imath\,k\,y}}{k^{\,2}} k_{\,i} \delta(p_- -k_-)\,
\int d^2 z_{\bot} e^{\,\imath\,z^i\,(p_i\,-\,k_i)}\,\int dz^+ e^{\,\imath\,z^+\,(p_+\,-\,k_+)}\tv^{\,cl}_{+}(z^+, z_{\bot}) \nonumber 
\eeqar
The second order contribution has the following form:
\beqar\label{CO2}
I_2 & = &- 2 g^2\,(-\imath)^2\,\int \frac{d^4 p}{(2\pi)^4} \frac{e^{-\,\imath\,p\,(x\,-\,z)}}{p^{\,2}}\frac{p_{\,i}}{p_{-}}\,
\int d^4 z\, \tv^{\,cl}_{+}(z)\,
\int \frac{d^4 p_1}{(2\pi)^4} \frac{p_{1\,-}\,e^{-\,\imath\,p_1\,(z\,-\,z_1)}}{p_{1}^{\,2}}\, \nonumber \\
&\,&
\int d^4 z_1\, 2 \tv^{\,cl}_{+}(z_1)\,
\int \frac{d^4 k}{(2\pi)^4} \frac{e^{-\,\imath\,k\,(z_1\,-\,y)}}{k^{\,2}}\,k_{\,i}\,.
\eeqar
In the Eikonal approximation we have
\beq\label{CO3}
p_{1}^{\,2}\approx\,2\,p_{1\,-}\,p_{1\,+}
\eeq
and correspondingly obtain:
\beqar\label{CO4}
I_2 & = &- 2 g^2\,(-\imath)^2\,
\int \frac{d^4 p}{(2\pi)^4} \frac{e^{-\,\imath\,p\,x\,}}{p^{\,2}}\frac{p_{\,i}}{p_{-}}\,
\int \frac{d^4 k}{(2\pi)^4} \frac{e^{\,\imath\,k\,y}}{k^{\,2}}\,k_{\,i}\,
\nonumber \\
&\,&
\int d^4 z\,e^{\,\imath\,z\,p}\,\tv^{\,cl}_{+}(z)\,
\int \frac{d^4 p_1}{(2\pi)^4} \frac{e^{-\imath\,p_1\,(z-z_1 )}}{p_{1\,+}}\,
\int d^4 z_1\,e^{-\imath\,z_{1}\,k}\,\tv^{\,cl}_{+}(z_1)\,.
\eeqar
Writing the integral
\beq\label{CO41}
\int \frac{d p_{1\,+}}{2 \pi}\, \frac{e^{-\imath\,p_1\,(z-z_1 )}}{p_{1\,+}\,+\varepsilon\,/\,2 p_{1 -}}\,=\,
-\imath\theta(p_{1\,-})\theta(z-z_1)+\imath\theta(-p_{1\,-})\theta(z_1-z)\,
\eeq
we will account further the $\theta(p_{1\,-})$ contribution only for the sake of simplicity.
Now, performing integrating on $p_{1}$, we obtain finally for \eq{CO4}:
\beqar\label{CO7}
I_2 & = & -2 g^2 (-\imath)^3\,2\pi\,
\int \frac{d^4 p}{(2\pi)^4} \frac{e^{-\,\imath\,p\,x\,}}{p^{\,2}}\frac{p_{\,i}}{p_{-}}\,
\int \frac{d^4 k}{(2\pi)^4} \frac{e^{\,\imath\,k\,y}}{k^{\,2}}\,k_{\,i}\,\delta(p_- -k_-) \nonumber \\
&\,&
\int d^2 z_{\bot}\,e^{\imath \,z^{i}\,(p_i - k_i)}\,
\int dz^+ \,\tv^{\,cl}_{+}(z^+ , z_{\bot})\,e^{\imath\,p_+\,z^+}\,
\int^{z^{+}}_{-\infty}\,d z_{1}^{+}\,\tv^{\,cl}_{+}(z_{1}^{+}, z_{\bot})\,e^{-\imath\,k_{+}\,z_{1}^{+}}\,
\eeqar
The third order contribution is the following one:
\beqar\label{CO8}
&\,& I_3 = 2 g^3\,(-\imath)^3\,
\int \frac{d^4 p}{(2\pi)^4} \frac{e^{-\,\imath\,p\,x\,}}{p^{\,2}}\frac{p_{\,i}}{p_{-}}\,
\int \frac{d^4 k}{(2\pi)^4} \frac{e^{\,\imath\,k\,y}}{k^{\,2}}\,k_{\,i}\,
\int d^4 z\,e^{\,\imath\,z\,(p-p_1)}\tv^{\,cl}_{+}(z)
\nonumber \\
&\,&
\int \frac{d^4 p_1}{(2\pi)^4} \frac{1}{p_{1\,+}}
\int d^4 z_1\,e^{\,\imath\,z_{1}\,(p_1-p_2)}\tv^{\,cl}_{+}(z_1)
\int \frac{d^4 p_2}{(2\pi)^4} \frac{1}{p_{2\,+}}
\int d^4 z_2\,e^{\,\imath\,z_{2} (p_2-k)}\tv^{\,cl}_{+}(z_2)\,
\eeqar
Using again \eq{CO41} we obtain:
\beqar\label{CO10}
I_3 & = & 2 g^3 (-\imath)^{5}\,2\pi\,
\int \frac{d^4 p}{(2\pi)^4} \frac{e^{-\,\imath\,p\,x\,}}{p^{\,2}}\frac{p_{\,i}}{p_{-}}\,
\int \frac{d^4 k}{(2\pi)^4} \frac{e^{\,\imath\,k\,y}}{k^{\,2}}\,k_{\,i}\,\delta(p_- -k_-) 
\int d^2 z_{\bot}\,e^{\imath \,z^{i}\,(p_i - k_i)}\,
\nonumber \\
&\,&
\int dz^+ \,\tv^{\,cl}_{+}(z^+ , z_{\bot})\,e^{\imath\,p_+\,z^+}\,
\int^{z^{+}}_{-\infty}\,d z^{+}_{1}\,\tv^{\,cl}_{+}(z^{+}_{1}, z_{\bot})\,
\int^{z^{+}_{1}}_{-\infty}\,d z^{+}_{2}\,\tv^{\,cl}_{+}(z^{+}_{2}, z_{\bot})\,
e^{-\imath\,k_{+}\,z^{+}_{2}}\,
\eeqar
The value of $p,k$ integrals in the expressions above are defined by poles 
\beq\label{CO81}
p_{+}\,=\,p_{\bot}^2/p_{-}\,,\,k_{+}\,=\,k_{\bot}^2/p_{-}\,,\,\,\,\,\,\,\,
p_{\bot}^2/p_{-}\,,\,k_{\bot}^2/p_{-}\,\propto\,|t|/\sqrt{s}
\eeq
that allows to leading order to take at $s\,\rightarrow\,\infty$:
\beq\label{CO82}
e^{\imath\,p_+\,z^+}\,\approx\,1\,,\,\,\,e^{-\imath\,k_{+}\,z^{+}_{i}}\,\approx\,\,\,1\,.
\eeq
We note, that this approximation is equivalent to the shock wave representation for the classical field when it takes as
$\tv_{+}(x^{+},\,x_{\bot})\,\propto\,\delta(x^{+})$, namely it will lead to the same result if we will substitute the classical fields in the form of the shock wave in the expression and 
will perform a re-summation of the perturbative terms. 
After the similar calculations of all order terms in \eq{Eff9} expression, we will obtain for the propagator of interests:
\beqar\label{CO11}
G_{+ +}^{ac}(x,y)\,& = &\,G_{0\,+ +}^{ac}(x,y)\,-\,4\pi\,\imath\,
\int \frac{d^4 p}{(2\pi)^4} \,e^{-\,\imath\,p\,x\,}
\int \frac{d^4 k}{(2\pi)^4} e^{\,\imath\,k\,y}\,\frac{1}{p^2\,k^2\,}\,
\frac{p_{i} k_{i}}{p_{-}}\,\nonumber \\
&\,&
\delta(p_- -k_-)\,
\int d^2 z_{\bot}\,e^{\imath \,z^{i}\, (p_i - k_i)}\,\Le\,\theta(p_{-})\, V^{a c}(\tv_{+}^{\,cl})\,-\,
\theta(-p_{-})\, V^{a c}(\tv_{+}^{\,cl})\,\Ra\,,
\eeqar
with
\beq\label{CO821}
V^{a b}_{\pm\,}(\tv_{\pm})\,=\,\Le P\,e^{g\,\int_{-\infty}^{\infty}\, d x^{\pm}\,\tv_{\pm}(x^{\pm},\,x^{\mp},\,x_{\bot})}\,-\,1\Ra^{a b}\,,
\eeq
for the adjoint representation of the gluon field.
This is a gluon propagator in an external field obtained in \cite{BalBel}, further we identify to LO precision $\tv_{+}^{\,cl}\,=\,\B_{+}$. 
Following to \cite{BalBel}, we also note that due the $\delta(p_- -k_-)$ in the expression, the integration on $p_{+}$ and $k_{+}$ variables provides
$\theta(x^{+})\,\theta(-y^{+})$ and $\theta(-x^{+})\,\theta(y^{+})$ in the answer, we will use this result in the derivation of BK equation.

\section{BK equation}
\label{BKE}

 In order to obtain the familiar form of BK equation from \eq{N221} expression, we consider the given terms of the \eq{N221} hierarchy and apply the shock wave approximation there.  
It means the following, 
taking any operator of interest in \eq{N221}, we have to define the expressions for these operators whereas the classical fields in the operators are distributed around $\delta(x^{+})$ only.
Therefore, we define:
\begin{equation}\label{BKE01}
\,O(\tv_{+})\,=\,
\left\{
\begin{array}{l l}
1\,&\,x^{+}\,<\,0 \\
U\,=\,P\,e^{g\,\int_{-\infty}^{\infty}\,d x^{+}\,\tv_{+}(x^{+},\,\,x_{\bot})}\,&\,x^{+}\,>\,0
\end{array}
\right.
\end{equation}
and
\begin{equation}\label{BKE011}
\,O^{T}(\tv_{+})\,=\,
\left\{
\begin{array}{l l}
U\,=\,P\,e^{g\,\int^{\infty}_{-\infty}\,d x^{+}\,\tv_{+}(x^{+},\,\,x_{\bot})}\,&\,x^{+}\,<\,0 \\
1\,&\,x^{+}\,>\,0 \,.
\end{array}
\right.
\end{equation}
Correspondingly, for the last term in \eq{N221} for example, we obtain:
\beqar\label{BKE012}
&\,&
\int^{\infty}_{-\infty}\,dx^+\,\Le O^{T}_{x}\,
(\imath\,T^{c}) \,O_{x}\,\Ra_{i k}\,
\int^{\infty}_{-\infty}\,dy^+\,\Le  O^{T}_{y}\,(\imath\,T^{d})
\,O_{y}\Ra_{l j}\,=\, \\
&=&
\int^{\infty}_{\varepsilon^{+}}dx^+ \Le 
(\imath T^{c}) U \Ra_{i k} 
\int^{-\varepsilon^{+}}_{-\infty} dy^+ \Le  U (\imath T^{d}) \Ra_{l j} + 
\int^{-\varepsilon^{+}}_{-\infty}\,dx^+ \Le U (\imath T^{c}) \Ra_{i k} 
\int^{\infty}_{\varepsilon^{+}} dy^+ \Le (\imath T^{d}) U\Ra_{l j} 
\eeqar
with the regularization of the integral around $x^{+}=0$ applied,
see also \eq{CO11} definition of the propagator which determines the separation of the integration regions.
Now, inserting the \eq{CO11} expression for the propagator and accountung that
\beq\label{BKE11}
<\ep_{\mu}(x)\,\ep_{\nu}(y)>\,=\,-\imath\,G_{\mu\,\nu}(x,y)\,.
\eeq
we obtain:
\beqar\label{BKE2}
&\,& J_1\,= \,g^2\,(- \imath)^2\,4\pi\,\int dx^{-}\,\delta(x^{-})\,\int dy^{-}\,\delta(y^{-})\,
\Le \Le (\imath T^{c}) U(x)\Ra_{i k}\,\Le U(y) (\imath T^{d})\Ra_{l j}\,+\,
\right.
\nonumber \\
&+&
\left.
\Le U(x) 
(\imath T^{c})\Ra_{i k}\,\Le (\imath T^{d}) U(y) \Ra_{l j}\Ra\, 
\int d x^+\,e^{-\imath\,x^{+}\,p_{+}}\,
\int d y^+\,e^{\imath\,y^{+}\,k_{+}}\,
\int \frac{d^4 p}{(2\pi)^4} \frac{e^{-\,\imath\,p_{-}\,x^{-}-\imath\,p_{i}\,x^{i}}}{p^{\,2}}\frac{p_{\,i}}{p_{-}}\,\theta(p_{-})
\nonumber \\
&\,&
\int \frac{d^4 k}{(2\pi)^4} \frac{e^{\imath\,k_{-}\,y^{-} + \imath\,k_{i}\,y^{i}}}{k^{\,2}} k_{\,i} \delta(p_- -k_-)\,
\int d^2 z_{\bot} e^{\,\imath\,z^i\,(p_i\,-\,k_i)}\,U^{c d}(z)\, =
\nonumber \\
&=&
-\,\frac{\alpha_s}{\pi^2}\,
\Le \Le (\imath T^{c}) U(x)\Ra_{i k}\,\Le U(y) (\imath T^{d})\Ra_{l j}\,+\,
\Le U(x) 
(\imath T^{c})\Ra_{i k}\,\Le (\imath T^{d}) U(y) \Ra_{l j}\Ra\, 
\int \frac{dp_{-}}{p_{-}}\,\theta(p_{-})\, 
\nonumber \\
&\,&
\int d^2 z_{\bot}\,
\frac{(x_{i} - z_{i})\,(y_{i} - z_{i})}{(z_{i} - x_{i})^2\,(y_{i} - z_{i})^2}\,
U^{c d}(z)\,,
\eeqar
here we used 
\beqar\label{BKE21}
\int d x^+\,e^{-\imath\,x^{+}\,p_{+}}\,=\,2\,\pi\,\delta(p_{+})\,\\
\int \frac{d^2 p_{i}}{(2\pi)^2} \frac{e^{-\imath\,p_{i}\,(z_{i} - x_{i})}\,p_i}{p_{i}^{2}-\imath\varepsilon}\,=\,
\frac{-\imath}{2\pi}\,\frac{z_{i} - x_{i}}{(z_{i} - x_{i})^2}
\eeqar
and considered only $\theta(p_{-})$ part of the propagator.

 An another contribution to BK equation we consider corresponds to the self-energy of the Wilson lines. 
Considering, for example, the fourth term in \eq{N221} we have there:
\beqar\label{BKE212}
&\,& J_2\,= \frac{g^2}{2}
\int^{\infty}_{-\infty}\,dx^+\,\int d^4 w\,\Le O^{T}_{x}\,(\imath\,T^{c})\,
G^{+}(x,w)\,(\imath\,T^{d})\,O_{w}\Ra_{ik} \,<\ep_{+}^{c}(x) \ep_{+}^{d}(w)>
\, =  
\nonumber \\
&=&
(-2\pi g^2)\,\int^{\infty}_{-\varepsilon^{+}}\,dx^+\,\int_{-\infty}^{\varepsilon^{+}} d w^{+}\,\int d^3 w\,
\Le (\imath\,T^{c})\,G^{+}(x,w)\,(\imath\,T^{d})\,\Ra_{ik}
\nonumber \\
&\,&
\int \frac{d^4 p}{(2\pi)^4} \,\frac{e^{-\,\imath\,p\,x\,}}{p^{\,2}}\frac{p_{\,i}}{p_{-}}\,
\theta(p_{-})\,\int \frac{d^4 k}{(2\pi)^4} \frac{e^{\,\imath\,k\,w}}{k^{\,2}}\, k_{\,i} \,\delta(p_- -k_-)\,
\int d^2 z_{\bot}\,e^{\imath \,z^{i}\, (p_i - k_i)}\, U^{c d}(\tv_{+}^{\,cl})\,.
\eeqar
Using \eq{N3} definition of $G^{+}(x,w)$ Green's function, we can write for the case of shock wave:
\beq\label{BKE21201}
\int^{\infty}_{0}\,dx^+\,\int_{-\infty}^{0} d w^{+}\,G^{+}(x,w)\,\propto\,
\int^{\infty}_{0}\,dx^+\,\int_{-\infty}^{0} d w^{+}\,P\,e^{g\int_{w^{+}}^{x^{+}}\,dx^{+}\,\tv_{+}\,d}\,=\,U(x_{\bot})\,
\int^{\infty}_{0}\,dx^+\,\int_{-\infty}^{0} d w^{+}\,
\eeq
and performing subsequent integration over $x^{+}$ and $w^{+}$, we will obtain for the \eq{BKE212} expression:
\beq\label{BKE125}
J_{2}^{1}\,=\,\frac{g^2}{8\pi^3}\,\Le (\imath\,T^{c}) U(\B_{+}) (\imath\,T^{d}) \Ra_{ik}\,
\int \frac{dp_{-}}{p_{-}}\, \int d^2 z_{\bot}\,
\frac{(x_{i} - z_{i})\,(x_{i} - z_{i})}{(z_{i} - x_{i})^2\,(x_{i} - z_{i})^2}\,
U^{c d}(B_{+})\,,
\eeq
Given
the fifth term in \eq{N22}, which is symmetrical with respect to the permutation of $x^+$ and $w^{+}$ variables, we obtain the following answer for these two terms and for two additional terms with $x\rightarrow\,y$ substitution: 
\beqar\label{BKE126}
&\,&J_{2} =\frac{g^2}{4\pi^3}
\Le (\imath\,T^{c}) U(\B_{+}) (\imath\,T^{d})\Ra_{ik}
\,V_{l j}(y_{\bot})\,
\int \frac{dp_{-}}{p_{-}}\, \int \frac{d^2 z_{\bot}}{(z_{i} - x_{i})^2}\,U^{c d}(B_{+})+
\nonumber \\
&+&
\frac{g^2}{8\pi^3}\,V_{i k}(x_{\bot}) \Le (\imath\,T^{c}) U(\B_{+}) (\imath\,T^{d})\Ra_{l j}
\int \frac{dp_{-}}{p_{-}}\, \int \frac{d^2 z_{\bot}}{(z_{i} - y_{i})^2}\,U^{c d}(B_{+})\,.
\eeqar
Therefore, we obtain finally for the correlator of interests:
\beqar\label{BKE127}
&\,& < \,V(x)\otimes  V(y) >=  V_{i k}(x_{\bot})\, V_{l j}(y_{\bot}) \,-\, 
\nonumber \\
&-&\,
\frac{\alpha_s}{\pi^2}\,
\Le \Le (\imath T^{c}) U(x)\Ra_{i k}\,\Le U(y) (\imath T^{d})\Ra_{l j}\,+\,
\Le U(x) 
(\imath T^{c})\Ra_{i k}\,\Le (\imath T^{d}) U(y) \Ra_{l j}\Ra\, 
\int \frac{dp_{-}}{p_{-}}\,
\nonumber \\
&\,&
\int d^2 z_{\bot}\,
\frac{(x_{i} - z_{i})\,(y_{i} - z_{i})}{(z_{i} - x_{i})^2\,(y_{i} - z_{i})^2}\,
U^{c d}(z)\,+\,
\nonumber \\
&+&
\frac{\alpha_s}{\pi^2}\,
\Le (\imath\,T^{c}) U(\B_{+}) (\imath\,T^{d}) \Ra_{ik}
\,V_{l j}(y_{\bot})\,\int \frac{dp_{-}}{p_{-}}\,
\int \frac{d^2 z_{\bot}}{(z_{i} - x_{i})^2}\,U^{c d}(B_{+})+
\nonumber \\
&+&
\frac{\alpha_s}{\pi^2}\,V_{i k}(x_{\bot}) \Le (\imath\,T^{d}) U(\B_{+}) (\imath\,T^{c})\Ra_{l j}\,\int \frac{dp_{-}}{p_{-}}\,
 \int \frac{d^2 z_{\bot}}{(z_{i} - y_{i})^2}\,U^{c d}(\B_{+})\,,
\eeqar
the expression is BK equation in Balitsky's formulation, see \cite{Bal}.

\section{Transverse fields sub-leading contributions to the gluon propagator}
\label{trn}

 The simplest corrections to the Balitsky hierarchy which we can account in the present framework it is a correction to the gluon's propagator in the external field. 
Namely, considering the \eq{Eff8}-\eq{Eff91} expressions 
for the propagator, we note that there are corrections to the propagator arise form the transverse fields contributions, see also \eq{Ef2}. In the Lipatov's approach these fields are not independent,
due the non-locality of the amplitudes the only fields of interests are Reggeons, the transverse fields are expressed through them, see \cite{Our1,Our2}. Therefore, proceeding with the 
\eq{Eff8} expression for the full propagator in the external field, we able to account the LO contribution of the transverse field into the propagator. In turn, the corrections will be expressed in terms of the longitudinal external field preserving the self-consistency of the hierarchy.

 Now, first of all, we consider the contribution to the propagators arising from the $M_{i j}$ part of Lagrangian proportional to the transverse fields:
\beq\label{trn1}
M_{i j}^{a c}\,=\,-
2 g f_{abc} \Le \delta_{ij} \,\tv_{k}^{b\,cl} \D_{k}  -
2\, \tv_{j}^{b\,cl} \D_{i}  + \tv_{i}^{b\,cl} \D_{j}\Ra\,,
\eeq
this vertex also contributes in $G_{++}$ and $G_{i +}$ propagators.
Nevertheless, we put an attention that for the part of $G_{++}$ propagator in the external field we have the contribution proportional to
\beq\label{trn2}
\,2g\,f^{b f d}\int\,d^4 z\,G_{0\,+ i}^{ab}(x,z)\,\Le \delta_{ij} \,\tv_{k}^{f\,cl} \D_{k}  -
2\, \tv_{j}^{f\,cl} \D_{i}  + \tv_{i}^{f\,cl} \D_{j}\Ra_{z}\,\, G_{j +}^{dc}(z,y)\,=\,0
\eeq
after the expression for the bare propagator \eq{Eff12} is accounted. For the $G_{i +}$ propagator the similar contributions gives
\beq\label{trn3}
\Le -(\tv^{b\,cl}_{i}\,p_{i})\,p_{j}\,+\,\tv^{b\,cl}_{j}\,(p_{i}\,p_{i})\Ra_{z}\,,
\eeq
which does not give zero  in \eq{Eff91} expression in general. Nevertheless, we know that the classical transverse field has the following structure:
\beq\label{trn4}
\tv_{i}^{b\,cl}\,=\,tr\,[f^{b}\,O(x^{+})\,
f^{c}\,O^{T}(x^{+})]\,\rho_{c i}\Le x^{-} , x_{\bot}\Ra \,=\,\frac{1}{N}\,
W^{b c}(x^{+})\rho_{c i}\Le x^{-} , x_{\bot}\Ra \,
\eeq
with
\beq\label{trn5}
\rho_{i}^{a}\,=\,
\Le\,\D_{-}^{-1}\,\Le\D_{i}\,\B_{-}^{b}\Ra\,\Ra\,,
\eeq
here $B_{-}(x^{-},\,x_{\bot})$ is the second Reggeon field. We see, that to this precision we can write this field in \eq{trn4} and correspondingly in \eq{trn3} as
\beq\label{trn6}
\textsl{v}_{i}^{a}\,=\,\frac{1}{N}\,W^{a b}\,\Le \textsl{v}_{+} \Ra\,
\Le\,\D_{-}^{-1}\,\B_{-}^{b}\,\Ra\,\D_{i}\,,
\eeq
that also provides zero contribution for the \eq{trn3} vertex in the external field part of \eq{Eff91} propagator.
Therefore, we conclude, that the transverse field part of $M_{i j}$ vertex does not contribute to our equation to this precision order

 An another contribution we have to discuss is contribution from $M_{- i}$ vertex to the corresponding propagators.
Whereas the system of \eq{Eff9}-\eq{Eff91} equations for $G_{++}$ and $G_{i +}$ full propagators can be resolved and written precisely in the form of infinite perturbative series,
we will use the truncated full solution for $G_{++}$ propagator. In shortened notations it reads as:
\beqar\label{trn7}
G_{++}\,& = &\,G_{0\,++}\,-\,G_{0\,++}\,M_{- j}\,G_{0\,j +}\,-\,\,G_{0\,+j}\,M_{j -}\,G_{0\,+ +}\,-\,
\nonumber \\
&-&
\,G_{0\,+j}\,M_{j l}\,\Le\, \delta_{l k}\,+\,G_{0\,l s}\,M_{s i}\,\Ra^{-1}\,\Le
G_{0\,i +}\,-\,G_{0\,i +}\,M_{- k}\,G_{0\,k +}\,-\,G_{0\,i p}\,M_{p -}\,G_{0\,+ +}\,\Ra\,.
\eeqar
This expression is reduced to the previous one by taking $M_{- i}\,=\,0$, the $\B_{-}$ field here is the smallness parameter, it's smallness is related to the asymmetry
of the scattering process. Our next observation is that the contribution to the propagator's external field part is coming from $p_{+}\,=\,0$ momentum, therefore
we can neglect all $G_{0\,++}$ terms in \eq{trn7}, see \eq{Eff9} expression. Therefore, we obtain:
\beq\label{trn8}
G_{++}\,= \,G_{0\,++}\,-\,
G_{0\,+j}\,M_{j l}\,\Le\, \delta_{l k}\,+\,G_{0\,l s}\,M_{s i}\,\Ra^{-1}\,\Le
G_{0\,i +}\,-\,G_{0\,i +}\,M_{- k}\,G_{0\,k +}\,\Ra\,.
\eeq
The last term in the \eq{trn7} expression determines the additional contributions to the BK equation provided by the classical $\textsl{v}_{i}$ gluon fields.
Correspondingly, repeating the \eq{BKE2} calculations, we obtain for the momentum dependent contribution to \eq{trn8}:
\beqar\label{trn9}
&\,& K_1\,= - \,g^3\,\int dx^{-}\,\delta(x^{-})\,\int dy^{-}\,\delta(y^{-})\,
\nonumber \\
&\,&
\int d x^+\,e^{-\imath\,x^{+}\,p_{+}}\,
\int d y^+\,(- \imath)^2\,4\pi\,
\int \frac{d^4 p}{(2\pi)^4} \frac{e^{-\,\imath\,p_{-}\,x^{-}-\imath\,p_{i}\,x^{i}}}{p^{\,2}}\frac{p_{\,i}}{p_{-}}
\nonumber \\
&\,&
\int \frac{d^4 k}{(2\pi)^4} \frac{e^{\imath\,k\,z_{1} }}{k^{\,2}} k_{\,i}\, \delta(p_- -k_-)\,
\int d^2 z_{\bot} e^{\,\imath\,z^i\,(p_i\,-\,k_i)}\,\int dz^+ e^{\,\imath\,z^+\,(p_+\,-\,k_+)}\,g\,
\tv^{\,cl\,c c_1}_{+}(z^+, z_{\bot}) 
\nonumber \\
&\,&
\int\,d^4\,z_{1}\,\Le \tv^{\,cl\,c_1 d}_{j}(z^{-}_{1}, z_{1\,\bot})\,\D_{-}\,-\,\Le \D_{-} \,\tv^{\,cl\,c_1 d}_{j}(z^{-}_{1}, z_{1\,\bot})\Ra\,\Ra\,
\int \frac{d^4 k_{1}}{(2\pi)^4} \frac{e^{-\imath\,k_{1\,}\,(z_{1}-y) }}{k^{\,2}}\frac{ k_{1\,j}}{k_{1\,-}}\,,
\eeqar
here, for the sake of simplicity, we preserved only the first perturbative contribution from the expression of the gluon's propagator. We also use 
\eq{trn4} expression for the transverse field at $\tv^{cl}_{+}\,=\,0\,$, the color structure of the expression stays the same as in LO BK equation.
Performing the simple integrations we will obtain:
\beqar\label{trn10}
&\,& K_1\,= - \,g^3\,(- \imath)^2\,4\pi\,(2\,\pi)^{2}\,
\int \frac{dp_{-}}{p_{-}}\,
\int \frac{d^2 p_{i}}{(2\pi)^4} \frac{e^{-\imath\,p_{i}\,x^{i}}\,p_i}{-p_{i}^{2}+\imath\varepsilon}\,
\nonumber \\
&\,&
\int \frac{d^4 k}{(2\pi)^4} \frac{ k_{\,i}\, \delta(p_- -k_-)}{k^{\,2}} \,
\int d^2 z_{\bot} e^{\,\imath\,z^i\,(p_i\,-\,k_i)}\,\int dz^+ e^{\,-\imath\,z^+\,k_+}\,g\,
\tv^{\,cl\,c c_1}_{+}(z^+, z_{\bot}) \,\int dz^{+}_{1}\,e^{\imath\,k_{+}\,z_{1}^{+}}\,
\nonumber \\
&\,&
\int d^2 z_{1\,\bot} e^{\imath\,z^{\,i}_{1}\, k_i}\,\int d z_{1\,-}\,e^{\imath\,z_{1}^{-}\,k_{-}}\,
\Le \tv^{\,cl\,c_1 d}_{j}(z^{-}_{1}, z_{1\,\bot})\,\D_{-}\,-\,\Le \D_{-} \,\tv^{\,cl\,c_1 d}_{j}(z^{-}_{1}, z_{1\,\bot})\Ra\,\Ra\,
e^{-\imath\,z_{1}^{-}\,k_{1\,-} }\,
\nonumber \\
&\,&
\int \frac{dk_{1\,-}}{k_{1\,-}}\,
\int \frac{d^2 k_{1\,i}}{(2\pi)^4} \frac{e^{-\imath\,k_{1\,i}\,(z_{1}^{\,i} - y^{i})}\,k_{1\,j}}{-k_{1\,i}^{2}+\imath\varepsilon}\,,
\eeqar
that in turn gives:
\beqar\label{trn11}
&\,& K_1\,= - \,g^3\,(- \imath)^2\,4\pi\,(2\,\pi)^{3}\,
\nonumber \\
&\,&
\int d^2 z_{\bot}\,\int d^2 z_{1\,\bot}\,
\int \frac{d^2 p_{i}}{(2\pi)^4} \frac{e^{-\imath\,p_{i}\,(x^{i} - z^{i})}\,p_i}{-p_{i}^{2}+\imath\varepsilon}\,
\int \frac{d^2 k_{i}}{(2\pi)^4} \frac{e^{-\imath\,k_{i}\,(z^{i} - z_{1}^{i})}\, k_{\,i}\, }{-k_{i}^{2}+\imath\varepsilon} \,
\nonumber \\
&\,&
\int \frac{d^2 k_{1\,i}}{(2\pi)^4} \frac{e^{-\imath\,k_{1\,i}\,(z_{1}^{\,i} - y^{i})}\,k_{1\,j}}{-k_{1\,i}^{2}+\imath\varepsilon}
\int dz^+ \,g\,
\tv^{\,cl\,c c_1}_{+}(z^+, z_{\bot}) \,
\int \frac{dp_{-}}{p_{-}}\,
\int \frac{dk_{1\,-}}{k_{1\,-}}\,
\nonumber \\
&\,&
\int d z_{1\,-}\,e^{\imath\,z_{1}^{-}\,p_{-}}\,
\Le \tv^{\,cl\,c_1 d}_{j}(z^{-}_{1}, z_{1\,\bot})\,\D_{-}\,-\,\Le \D_{-} \,\tv^{\,cl\,c_1 d}_{j}(z^{-}_{1}, z_{1\,\bot})\Ra\,\Ra\,
e^{-\imath\,z_{1}^{-}\,k_{1\,-} }\,.
\eeqar
We take into account now that for $\D_{-} \,\tv^{\,cl\,c_1 d}_{j}(z^{-}_{1}, z_{1\,\bot})$ term's contribution we obtain:
\beq\label{trn12}
\int \frac{dp_{-}}{p_{-}}\,
\int \frac{dk_{1\,-}}{k_{1\,-}}\,e^{-\imath\,z_{1}^{-}\,(k_{1\,-} - p_{-}) }\propto\,\theta(z_{1})\,\theta(-z_{1})\,=\,0
\eeq
and, therefore, we have:
\beqar\label{trn1201}
&\,& K_1\,= - \,g^3\,(- \imath)^3\,4\pi\,(2\,\pi)^{3}\,
\int \frac{dp_{-}}{p_{-}}\,
\nonumber \\
&\,&
\int d^2 z_{\bot}\,\int d^2 z_{1\,\bot}\,
\int \frac{d^2 p_{i}}{(2\pi)^4} \frac{e^{-\imath\,p_{i}\,(x^{i} - z^{i})}\,p_i}{-p_{i}^{2}+\imath\varepsilon}\,
\int \frac{d^2 k_{i}}{(2\pi)^4} \frac{e^{-\imath\,k_{i}\,(z^{i} - z_{1}^{i})}\, k_{\,i}\, }{-k_{i}^{2}+\imath\varepsilon} \,
\nonumber \\
&\,&
\int \frac{d^2 k_{1\,i}}{(2\pi)^4} \frac{e^{-\imath\,k_{1\,i}\,(z_{1}^{\,i} - y^{i})}\,k_{1\,j}}{-k_{1\,i}^{2}+\imath\varepsilon}
\int dz^+ \,g\,
\tv^{\,cl\,c c_1}_{+}(z^+, z_{\bot}) \,
\int d z_{1\,-}\,
\tv^{\,cl\,c_1 d}_{j}(z^{-}_{1}, z_{1\,\bot})\,\delta(z^{-}_{1})\,.
\eeqar
Now, we obtain the following term additionally to \eq{BKE2} answer:
\beqar\label{trn13}
K_1  & = &
\frac{\alpha_s}{\pi^2}\,\Le \Le (\imath T^{c}) U(x)\Ra_{i k}\,\Le U(y) (\imath T^{d})\Ra_{l j}\,+\,
\Le U(x) 
(\imath T^{c})\Ra_{i k}\,\Le (\imath T^{d}) U(y) \Ra_{l j}\Ra\, 
\nonumber \\
&\,&
\int \frac{dp_{-}}{p_{-}}\,\theta(p_{-})\, 
\int d^2 z_{\bot}\,\int d^2 z_{1\,\bot}\,U^{c c_1}(z_{\bot})\,
\frac{(x_{i} - z_{i})\,(z_{i} - z_{1\,i})(z_{1,j} - y_{j})}{(z_{i} - x_{i})^2\,(z_{i} - z_{1\,i})^2\,(y_{i} - z_{1\,i})^2}\, \tilde{\rho}_{j}^{c_1 d}(z_{1\,\bot})
\eeqar
with
\beq\label{trn14}
\tilde{\rho}_{j}^{c_1 d}(z_{1\,\bot})\,=\,2 \pi g\,\int d z_{1\,-}\,
\tv^{\,cl\,c_1 d}_{j}(z^{-}_{1}, z_{1\,\bot})\,\delta(z^{-}_{1})\,.
\eeq
The \eq{trn14} expression, in turn, depends on the form of $\D_{-}^{-1}$ operator which appears in \eq{trn6}. Namely, taking for example
\beq\label{trn15}
\D_{-}^{-1}\,=\,\frac{1}{2}\,\Le \int_{-\infty}^{z^{-}_{1}} d w^{-}
\,-\,\int^{\infty}_{z^{-}_{1}} d w^{-} \Ra
\eeq
we will obtain for \eq{trn14} expression with the use of LO value of the transverse field from \eq{trn4}-\eq{trn6}:
\beq\label{trn16}
\tilde{\rho}_{j}^{c_1 d}(z_{1\,\bot})\,=\,\pi\, g\,f^{a c_1 d}\,\Le\int^{0}_{-\infty} d w^{-}\,
 \D_{j}\,\B^{\,a}_{-}(w^{-},\,z_{1\,\bot})\,-\,\int_{0}^{\infty} d w^{-}\,
 \D_{j}\,\B^{\,a}_{-}(w^{-},\,z_{1\,\bot})\Ra\,,
\eeq
here the integral's value depends on the parity of the physical source (impact factor) of $\B_{-}$ Reggeon field.
In the case of the shock wave approximation when $\B_{-}\,\propto\,\delta(w^{-})$ the \eq{trn16} is zero of course, there the non-zero contribution will
appear only from the correction to the $\delta(w^{-})$ distribution of the classical external field. We also note, that for the shock wave, the \eq{trn4}
expression can be rewritten as
\beq\label{trn161}
\frac{1}{N}\,W^{a b}(\B_{+})\,\rightarrow\,\delta^{a b}\,U(B_{+})\,
\eeq
which provides, for example, additional correcton for the $\D^{-1}_{-}$ operator different from \eq{trn15}.

  Performing the similar calculations with respect to \eq{BKE125} contributions to BK equation, we obtain:
\beqar\label{trn17}
K_{2}^{1}\,& = &\,-\,\frac{\alpha_s}{\pi^2} \,\Le (\imath\,T^{c}) V(\B_{+}) (\imath\,T^{d})\Ra_{ik}
\,U_{l j}(y_{\bot})\,
\nonumber \\
&\,&
\int \frac{dp_{-}}{p_{-}}\,
\int d^2 z_{\bot}\,\int d^2 z_{1\,\bot}\,U^{c c_1}(z_{\bot})\,
\frac{(x_{i} - z_{i})\,(z_{i} - z_{1\,i})(z_{1,j} - x_{j})}{(z_{i} - x_{i})^2\,(z_{i} - z_{1\,i})^2\,(x_{i} - z_{1\,i})^2}\, \tilde{\rho}_{j}^{c_1 d}(z_{1\,\bot})\,.
\eeqar
Therefore, the NLO sub-leading contributions to the BK equation arise from the classical transverse field contribution have the following form:
\beqar\label{trn18}
&\,& < T^{a}\,V(x)\otimes T^{b} V(y) >_{Tr}= \, \nonumber \\
&=&
\frac{\alpha_s}{\pi^2}\,\Le \Le (\imath T^{c}) U(x)\Ra_{i k}\,\Le U(y) (\imath T^{d})\Ra_{l j}\,+\,
\Le V(x) 
(\imath T^{c})\Ra_{i k}\,\Le (\imath T^{d}) V(y) \Ra_{l j}\Ra\, 
\nonumber \\
&\,&
\int \frac{dp_{-}}{p_{-}}\, 
\int d^2 z_{\bot}\,\int d^2 z_{1\,\bot}\,U^{c c_1}(z_{\bot})\,
\frac{(x_{i} - z_{i})\,(z_{i} - z_{1\,i})(z_{1,j} - y_{j})}{(z_{i} - x_{i})^2\,(z_{i} - z_{1\,i})^2\,(y_{i} - z_{1\,i})^2}\, \tilde{\rho}_{j}^{c_1 d}(z_{1\,\bot})\,-
\nonumber \\
&-&
\frac{\alpha_s}{\pi^2} \,\Le (\imath\,T^{c}) U(\B_{+}) (\imath\,T^{d})\Ra_{ik}\,V_{l j}(y_{\bot})\,
\nonumber \\
&\,&
\int \frac{dp_{-}}{p_{-}}\,
\int d^2 z_{\bot}\,\int d^2 z_{1\,\bot}\,U^{c c_1}(z_{\bot})\,
\frac{(x_{i} - z_{i})\,(z_{i} - z_{1\,i})(z_{1,j} - x_{j})}{(z_{i} - x_{i})^2\,(z_{i} - z_{1\,i})^2\,(x_{i} - z_{1\,i})^2}\, \tilde{\rho}_{j}^{c_1 d}(z_{1\,\bot})\,-
\nonumber \\
&-&
\frac{\alpha_s}{\pi^2} V_{i k}(x_{\bot}) \Le (\imath\,T^{d}) U(\B_{+}) (\imath\,T^{c})\Ra_{l j}\,
\nonumber \\
&\,&
\int \frac{dp_{-}}{p_{-}}\,
\int d^2 z_{\bot}\,\int d^2 z_{1\,\bot}\,U^{c c_1}(z_{\bot})\,
\frac{(y_{i} - z_{i})\,(z_{i} - z_{1\,i})(z_{1,j} - y_{j})}{(z_{i} - y_{i})^2\,(z_{i} - z_{1\,i})^2\,(y_{i} - z_{1\,i})^2}\, \tilde{\rho}_{j}^{c_1 d}(z_{1\,\bot})\,.
\eeqar
We see, that this correction is of the eikonal type and its contribution crucially depend on the form of \eq{trn15} operator. Taking it symmetrical for the case of the shock wave approximation, the contribution will not be zero only when 
the corrections to the $\delta(x^{\pm})$ shock wave form of the classical fields will be accounted. In this extend we can consider these corrections as corrections beyond shock wave approximations.

\section{Sub-leading eikonal corrections to the quark propagator}
\label{NonEik}

 As mentioned above, we can consider the Wilson line in \eq{WL} Lipatov's action as first order approximation to the \eq{SSec23} "phase" operator \eq{SSec23},
see for example also \cite{Meggio,Laenen,Chiril}.
This is important observation, because it allows to extend the initial formulation of the effective action beyond the eikonal approximation and additionally allows to account the sub-leading eikonal corrections to the ordinary Wilson line. An another important lesson is that going to account these corrections, we need to write in the action separately two covariant currents with Reggeons as sources:
the first one for the fundamental Wilson line and the second one for the adjoint. As discussed, the reason for that is the coincidence of the forms of these Wilson line only to LO precision, in general
they are different. We consider further the case of the fundamental Wilson lines, for the adjoint ordered exponential the calculations can be done similarly of course, with only change
of the \eq{SSec23} "phase" operator on the gluon's one.

  Therefore, going beyond the eikonal approximation, we can use the following expression
instead \eq{WL2101} form of the operator:
\beqar\label{NE01}
&\,&V(\tv)\,=\,f(\infty, -\infty)\,=\,\mathcal{N}^{-1}\,
\\
&\,&
\int \mathcal{D} \xi
e^{\frac{\imath}{2}\int_{-\infty}^{\infty} dt \dot{\xi}^2}
\Le P Exp\left[\imath\int_{x_{i}(-\infty)}^{x_{f}(\infty)} dx^{\mu} \tv_{\,\mu}(x)+\imath\int_{-\infty}^{\infty} dt \Le
\dot{\xi}^{\mu} \tv_{\,\mu}(x) +\frac{1}{4}\sigma^{\mu \nu} F_{\mu \nu}(x)\Ra \right] - 1\Ra \,,
\nonumber 
\eeqar
with \eq{SSec191} as a parametrization of the trajectory of the quark\footnote{For an arbitrary kinematic and configuration of the gluon fields the
equations of motion for the $x$ can be solved firstly that determines a calculation scheme for the path integral in respect to the fluctuations $\xi$.} and
the following normalization of the $\xi$ path integral in \eq{NE01}:
\beq\label{NE4}
\mathcal{N}(\infty, -\infty)\,=\,\int\,\mathcal{D} \xi\,Exp\left[\,\frac{\imath}{2}\,\int_{-\infty}^{\infty}\,dt\,  \dot{\xi}^2\,\right]\,.
\eeq
This normalization, as we mentioned above, corresponds to the definition of the truncation of the quark's propagator required in the Lipatov's formalism.
In our case, when we talk about Regge kinematics, 
we have:
\beq\label{NE03}
x\,=\,p_{f}\,t\,+\,\xi(\lambda)\,,\,\,\,p^{\,\mu}_{f}\,=\,\sqrt{\frac{s}{2}}\,n_{LC}^{\mu}\,,\,\,\,n_{LC}^{\mu}=(1,0,0_{\bot})\,,\,\,\,n_{LC}^{2}\,=\,0\,
\eeq  
and it provides:
\beq\label{NE1}
V(\tv)\,=\,P\,\left[\Le\, e^{g\,\int_{-\infty}^{\infty}\,dx^{\mu}\,\tv_{\mu}\,+\,
\frac{g}{2\,\sqrt{s}}\,\int_{-\infty}^{\infty}\,d\lambda\,\sigma^{\mu\,\nu}\,F_{\mu\,\nu}}\Ra_{\xi\,=\,0}\,S(\xi)\,-\,1\right]\,,
\eeq 
with the following definitions of the quark's trajectory defined:
\beq\label{NE2}
x\,=\,\sqrt{\frac{s}{2}}\,n_{LC}\,t\,+\,\xi(\lambda)\,=\,\frac{\lambda}{\sqrt{2}}\,n_{LC}\,+\,\xi(\lambda)\,,\,\,\,\xi(\infty)\,=\,\xi(-\infty)\,=\,0\,.
\eeq
The variable $\xi$ in the expressions is a fluctuation of the  trajectory around the straight line, with $F_{\mu\nu}$ as the gluon field strength tensor we have:
\beqar\label{NE3}
&\,& S(\xi)\,=\,\mathcal{N}^{-1}\,\int\,\mathcal{D} \xi\,Exp\left[\,\imath\,\int\,d\lambda\,\Le\, \frac{\,\sqrt{s}}{2}\, \dot{\xi}^2 -\imath\,g\,\dot{\xi}^{\mu}\,\sum_{n=1}^{\infty}
C(n)\,\tv_{\,\mu,\,\rho_1\cdots\rho_n}\,\xi^{\rho_1}\cdots\xi^{\rho_n}\,-\,
\right.\right.
\nonumber \\
&-&\left.\left.
\frac{\imath\,g}{2\,\sqrt{s}}\, \sigma^{\mu\nu}
\sum_{n=1}^{\infty} C(n) F_{\mu \nu ,\,\rho_1\cdots\rho_n}\, \xi^{\rho_1}\cdots\xi^{\rho_n}
\Ra\right].
\eeqar
The \eq{NE3} $S(\xi)$ factor in \eq{NE1}
accounts the sub-leading non-eikonal corrections which are related to the fluctuations around the straight line trajectory. Namely, we
can solve equations of motion for the $\xi$ and calculate the corresponding contributions of the \eq{NE3} path integral into \eq{NE1} Wilson line, see for example \cite{Chiril,Arm}.
Further we do not consider these types of the 
non-eikonal contributions taking this factor equal to one, the calculations of it's contribution to the BK equation we will perform in an additional publication.

 In our case, therefore, we have the following expression for the eikonal ordered exponential with sub-leading corrections included:
\beq\label{NE5}
V(\tv)\,=\,P\, e^{\,g\,\int_{-\infty}^{\infty}\,dx^{\mu}\,\tv_{\mu}\,+\,
\frac{\,g}{2\,\sqrt{s}}\,\int_{-\infty}^{\infty}\,d\lambda\,\sigma^{\mu\,\nu}\,F_{\mu\,\nu}}\,-\,1\,.
\eeq 
Taking into account the contribution of $\tv_{+}$ gluon field in \eq{NE5} expression
and adjusting \eq{NE2} definitions
\beq\label{NE51}
\lambda\,=\,\sqrt{2}\,\tilde{n}_{\mu}\,x^{\mu}\,,\,\,\,\tilde{n}_{\mu}\,=\,(1,0,0_{\bot})\,,
\eeq
we obtain correspondingly:
\beq\label{NE6}
V(\tv)\,=\,P\, e^{\,g\,\int_{-\infty}^{\infty}\,dx^{+}\,\tv_{+}\,+\,
g\,\sqrt{\frac{2}{s}}\,\int_{-\infty}^{\infty}\,dx^{+}\,\tilde{n}_{+}\,\Le\, \sigma^{-\,+}\,\D_{-}\tv_{+} + \sigma^{i\,+}\D_{i}\tv_{+}\Ra}\,-\,1\,.
\eeq 
Expanding the Wilson line with respect to the fluctuations around the classical Reggeon, see \eq{N1}, we will obtain instead \eq{N2}:
\beqar\label{NE7}
V(\tv_{+})\,& = &\,V(\B_{+}(x^{+},x_{\bot}))\,+\,
\nonumber \\
&+&
g\,\int^{\infty}_{-\infty}\,dx^+ O^{T}(\B_{+}(x^{+},x_{\bot}))
\Le \ep_{+}(x)\,+\,\sqrt{\frac{2}{s}}\,\tilde{n}_{+}\,\Le\, \sigma^{-\,+}\,\D_{-}\ep_{+} + \sigma^{i\,+}\D_{i}\ep_{+}\Ra\Ra
O(\B_{+}(x^{+},x_{\bot})) +\,\nonumber\\
&+&\,
\frac{g^2}{2}\,\int^{\infty}_{-\infty}\,dx^+\,O^{T}(\B_{+}(x^{+},x_{\bot}))\,
\Le \ep_{+}(x)\,+\,\sqrt{\frac{2}{s}}\,\tilde{n}_{+}\,\Le\, \sigma^{-\,+}\,\D_{-}\ep_{+} + \sigma^{i\,+}\D_{i}\ep_{+}\Ra\Ra
\,\nonumber \\
&\,&
\,\int d^4 p\,
G^+\,(x,p)\,
\Le \ep_{+}(p)\,+\,\sqrt{\frac{2}{s}}\,\tilde{n}_{+}\,\Le\, \sigma^{-\,+}\,\D_{-}\ep_{+} + \sigma^{i\,+}\D_{i}\ep_{+}\Ra\Ra
\,O(\B_{+}(p^{+},p_{\bot}))\,+\,\nonumber \\
&+&\,
\frac{g^2}{2}\,\int^{\infty}_{-\infty}\,dx^+\,\int d^4 p\,O^{T}\,(\B_{+}(p^{+},p_{\bot}))\,
\Le \ep_{+}(p)\,+\,\sqrt{\frac{2}{s}}\,\tilde{n}_{+}\,\Le\, \sigma^{-\,+}\,\D_{-}\ep_{+} + \sigma^{i\,+}\D_{i}\ep_{+}\Ra\Ra
\nonumber \\
&\,&
 G^+\,(p,x)\,
\Le \ep_{+}(x)\,+\,\sqrt{\frac{2}{s}}\,\tilde{n}_{+}\,\Le\, \sigma^{-\,+}\,\D_{-}\ep_{+} + \sigma^{i\,+}\D_{i}\ep_{+}\Ra\Ra
\,O(\B_{+}(x^{+},x_{\bot}))\,+\cdots,
\eeqar
with
\beq\label{NE8}
V(\B_{+})\,=\,P\, e^{\,g\,\int_{-\infty}^{\infty}\,dx^{+}\,\B_{+}\,+\,
g\,\sqrt{\frac{2}{s}}\,\int_{-\infty}^{\infty}\,dx^{+}\,\tilde{n}_{+}\,\sigma^{i\,+}\D_{i}\B_{+}}\,-1\,,\,\,\,\D_{-}\B_{+}\,=\,0\,
\eeq 
definition used. Now, correspondingly, we obtain instead \eq{N22}:
\beqar\label{NE9}
&\,& < T^{a}\,V(x)\otimes T^{b} V(y) >= T^{a} V(x_{\bot})\,T^{b} V(y_{\bot})\,+\,
\nonumber \\
&+&
g^2 
\int^{\infty}_{-\infty}\,dx^+\,\Le T^{a}\,O^{T}_{x}\,
(\imath\,T^{c}) \,O_{x}\,\Ra\,
\int^{\infty}_{-\infty}\,dy^+\,\Le T^{b}\,O^{T}_{y}\,(\imath\,T^{d})
\,O_{y}\Ra\,
\nonumber \\
&\,&
\Le 1 + \sqrt{\frac{2}{s}}\,\tilde{n}_{+}\,\Le\, \sigma^{-\,+}\,\Le \D_{x\,-} +\D_{y\,-}\Ra + \sigma^{i\,+}\Le \D_{x\,i} + \D_{y\,i}\Ra\Ra\Ra\,
< \ep_{+}^{c}(x) \ep_{+}^{d}(y) >\,+\, 
\nonumber \\
&+&
\frac{g^2}{2}
\int^{\infty}_{-\infty}\,dx^+\,\int d^4 p\,\Le T^{a}\,O^{T}_{x}\,(\imath\,T^{c})\,
G^{+}(x,p)\,(\imath\,T^{d})\,O_{p}\Ra \,
\nonumber \\
&\,&
\Le 1 + \sqrt{\frac{2}{s}}\,\tilde{n}_{+}\,\Le\, \sigma^{-\,+}\,\Le \D_{x\,-} +\D_{p\,-}\Ra + \sigma^{i\,+}\Le \D_{x\,i} + \D_{p\,i}\Ra\Ra\Ra\,
<\ep_{+}^{c}(x) \ep_{+}^{d}(p)>
T^{b}\, V(y_{\bot}) +  \nonumber \\
&+&
\frac{g^2}{2}
\int^{\infty}_{-\infty}\,dx^+\,\int d^4 p\,\Le T^{a}\,O^{T}_{p}\,(\imath\,T^{c})\,
G^{+}(p,x)\,(\imath\,T^{d})\,O_{x}\Ra \,
\nonumber \\
&\,&
\Le 1 + \sqrt{\frac{2}{s}}\,\tilde{n}_{+}\,\Le\, \sigma^{-\,+}\,\Le \D_{x\,-} +\D_{p\,-}\Ra + \sigma^{i\,+}\Le \D_{x\,i} + \D_{p\,i}\Ra\Ra\Ra\,
<\ep_{+}^{c}(p) \ep_{+}^{d}(x)>
T^{b}\, V(y_{\bot}) +  \nonumber \\
&+&
\frac{g^2}{2}T^{a}\, V(x_{\bot})\,
\int^{\infty}_{-\infty}\,dy^+\,\int d^4 p\,\Le T^{b}\,O^{T}_{p}\,(\imath\,T^{c})\,
 G^+\,(p,y)\,(\imath\,T^{d})\,O_{y}\Ra\,
\nonumber \\
&\,&
\Le 1 + \sqrt{\frac{2}{s}}\,\tilde{n}_{+}\,\Le\, \sigma^{-\,+}\,\Le \D_{y\,-} +\D_{p\,-}\Ra + \sigma^{i\,+}\Le \D_{y\,i} + \D_{p\,i}\Ra\Ra\Ra\,
<\ep_{+}^{c}(p) \ep_{+}^{d}(y)> +\nonumber \\
&+&
\frac{g^2}{2}T^{a}\, V(x_{\bot})\,
\int^{\infty}_{-\infty}\,dy^+\,\int d^4 p\,\Le T^{b}\,O^{T}_{y}\,(\imath\,T^{c})\,
 G^+\,(y,p)\,(\imath\,T^{d})\,O_{p}\Ra\,
\nonumber \\
&\,&
\Le 1 + \sqrt{\frac{2}{s}}\,\tilde{n}_{+}\,\Le\, \sigma^{-\,+}\,\Le \D_{y\,-} +\D_{p\,-}\Ra + \sigma^{i\,+}\Le \D_{y\,i} + \D_{p\,i}\Ra\Ra\Ra\,
<\ep_{+}^{c}(y) \ep_{+}^{d}(p)> \,.
\eeqar
We note, that due the $\delta(p_- - k_-)$ function present in the gluon's propagator, the contribution which arises from the action of the
$\D_{x\,-} +\D_{y\,-}$  operator on the propagator is zero. The same we obtain for the action of $\D_{x\,i} + \D_{y\,i}$ operator on the terms which represents the
Wilson's line self-energy leading order contribution, there $\delta^{2}(x_{\bot} - w_{\bot})$ function is present.
Therefore, we stay with the only following additional contribution to BK equation:
\beqar\label{NE901}
&\,&\Le \D_{x\,j} + \D_{y\,j}\Ra\,\frac{(x_{i} - z_{i})\,(y_{i} - z_{i})}{(x_{i} - z_{i})^2\,(y_{i} - z_{i})^2}\,=\,
\frac{1}{(x_{i} - z_{i})^2\,(y_{i} - z_{i})^2}\,\left[(y_j -z _j)\,+\,(x_j -z _j)\,-\,
\right.\nonumber \\
&-&
\left.
\frac{2\,(x_j -z _j)\,(x_{i} - z_{i})\,(y_{i} - z_{i})}{(x_{i} - z_{i})^2}\,-\,\frac{2\,(y_j -z _j)\,(x_{i} - z_{i})\,(y_{i} - z_{i})}{(y_{i} - z_{i})^2}\,\right]\,=\,
\nonumber \\
&=&
\frac{(y_j -z _j)\,(y_{i} - z_{i})\,(y_{i} - 2x_{i} +z_{i})}{(x_{i} - z_{i})^2\,(y_{i} - z_{i})^4}\,+\,
\frac{(x_j -z _j)\,(x_{i} - z_{i})\,(x_{i} - 2y_{i} +z_{i})}{(x_{i} - z_{i})^4\,(y_{i} - z_{i})^2}\,
\eeqar 
which we finally write in the full form as
\beqar\label{NE10}
&\,& < V(x)\otimes  V(y) >_{Q}= \, 
-\,\frac{\alpha_s}{\pi^2}\sqrt{\frac{2}{s}}\,
\nonumber \\
&\,&
\Le \Le (\imath T^{c}) U(x)\Ra_{i k}\,\Le U(y) (\imath T^{d})\Ra_{l j}\,+\,
\Le V(x) (\imath T^{c})\Ra_{i k}\,\Le (\imath T^{d}) U(y) \Ra_{l j}\Ra\,
\sigma^{j\,+}\,\int \frac{dp_{-}}{p_{-}}\,\int d^{\,2} z_{\bot}\, 
\nonumber \\
&\,&
\Le
\frac{(y_j -z _j)(y_{i} - z_{i})(y_{i} - 2x_{i} +z_{i})}{(x_{i} - z_{i})^2 (y_{i} - z_{i})^4}+
\frac{(x_j -z _j)(x_{i} - z_{i})(x_{i} - 2y_{i} +z_{i})}{(x_{i} - z_{i})^4 (y_{i} - z_{i})^2}
\Ra
U^{c d}(B_{+}).
\eeqar
This sub-leading correction is as well of the eikonal type, it describes the quark's spin contribution to BK equation in the processes without helicity flip.

\section{Conclusion}

 In this paper we demonstrated the interconnection between the Lipatov's QCD effective action and Balitsky description of the correlators of ordered exponentials
of classical external fields. 
The main result of the paper is that we show how  the correlators can be obtained directly from the effective action of Lipatov.

There are the following important issues that  we addressed in the course of the derivation.
The effective currents in the action of Lipatov and correlators of the approach of Balitsky are energetic quark's (or gluon's) propagators in an external field. These propagators, full or truncated, in the leading order  approximation can be written in terms of ordered exponentials. 
Therefore, there are two important issues we address: a) in which form and b) at what precision one must account for  the propagators. 
As it was already mentioned in \cite{Feynman}, the usual four dimensional spinor form of the quark's propagator is not convenient to use, consequently we consider two-dimensional form of the propagator which is easier to use in the processes without flip of the quark helicity. This simplification of the propagator's form makes the calculations simpler, in turn, the form of the propagators clarifies the corrections to the BK equation from both the spin's correction to the propagator and from the deviation of the trajectory of the quarks from the straight line, see \eq{NE01} and \eq{NE1}. Once  both corrections are written in the single expression one can  develop a perturbative expansion of the propagators with  respect to the fluctuations of the trajectory about the straight line.
Therefore, taking the effective current in the Lipatov's action in the  form of \eq{NE01}, we introduce  these type of the corrections in our approach  in a straightforward manner, that extend the 
area of the application of the Lipatov's effective action in both original, \cite{LipatovEff}, and new, \cite{OurZub}, forms.
Consequently, there is a new and interesting task to formulate the original version of the action, \cite{LipatovEff}, in terms of the Reggeon fields with included in it the \eq{NE1} operator. 
This action's generalization will provide the sub-leading and non-eikonal corrections to the amplitudes of the processes with Reggeon fields involved, see also \cite{Our5}.
We also conclude, that considering gluon and quark propagations both together in the action, we will need to add in the effective action, additionally to the \eq{NE1} expression, the similar operator arises from the propagation of the gluon in the external field, we plan to consider this task in an additional publication. 
The processes with the helicity flip is an another example where this extended action can be used. Namely, these processes
can be considered similarly to done with the only difference that we will need to use there the propagators
corresponding to the flips which can be introduced similarly to discussed in Section \ref{SS}.

 It is worth emphasizing that the Lipatov's action does not originally imply  the shock wave form of the external fields. The action is written for the Reggeon degrees of freedom, which interaction
is averaged with respect to the gluon and quark fields in some rapidity interval. In this case, the whole amplitude of any scattering process can be written as non-local interactions of these clusters by the reggeized gluons, the interactions of the Reggeons with the target and projectile at the edges of the interval determine the form of the Reggeon fields. In this extend, 
we obtained a precise system of equations for the ordered exponentials, similar to the \cite{Bal} hierarchy, directly from the action of Lipatov.
The new hierarchy is precise in the sense that it includes the ordered exponentials written in terms of general external fields beyond the shock wave approximation. The \eq{N221} equation can be written fully in terms of $O$ and $O^{T}$ operators, nevertheless it is not clear if we can resolve the equation in terms of the operators in this case.
Therefore, in order to obtain the familiar form of the BK equation it is crucial to assume the shock wave form for the external fields and after that one can write the equation in terms of the infinite Wilson line only. This different approach to the problem perhaps allows to introduce the corrections to the equation on some clear base. Namely, in the proposed approach we firstly write the equation and only after that we can expand any object in the equation with respect to any parameters of the problem: it can be finite range of the external field, corrections to the gluon's propagator,
non-straight line of the particle's trajectory, all possible corrections together and etc. 

 The central  result of the proposed framework is that it provides a useful instrument for the further perturbative calculations related to the BK equation.
First of all, the effective action allows to formulate the problem for the different QCD gauges directly in the action and, in turn, allows to determine corresponding redefinition of the propagators~(correlators) of interest which appear in r.h.s. of \eq{N221}. Additionally, it clarifies the translation vocabulary between the different approaches. For example, the variation of the action with respect to the auxiliary currents, \eq{WL21}, correspond to the correlators of the reggeized gluons fields in the complimentary theory written in terms of the Reggeons fields, there the reggeized gluons are sources of the curents, see \cite{Our5} for example.
From that it is clear, that the \eq{N2} expansion of the Wilson line is in correspondence with the trajectory of the reggeized gluon, the correlator of two Wilson lines, \eq{N22}, is in correspondence to the BFKL Pomeron, the correlators of the higher orders are in correspondence with the correlators written in terms of the $\B_{+}$ reggeized gluon fields. It is interesting to note, as it follows  from this relation, that the calculation of the $<B_{+}\,B_{+}>$ correlator in the original theory of the effective action, \cite{LipatovEff,Our1,Our2,Our3,Our4,Our5} with 
$\B_{-}\,\neq\,0,\,\B_{+}\,=\,0$ taken at the end and corresponding calculation of  $<B_{-}\,B_{-}>$  with $\B_{-}\,=\,0,\,\B_{+}\,\neq\,0$ at the end, will determine the forms of both
triple Pomeron vertices in the light-cone gauge, see effective theory for the Pomerons interactions with these vertices in \cite{BraBo}. It is important to mention, that because the 
\cite{LipatovEff}
effective action can be written in any gauge, the corresponding calculations can be done symmetrically with respect to target and projectile, i.e. symmetrically with respect to
Reggeon fields $\B_{-}$ and $B_{+}$ or symmetrically with respect to Wilson lines which depend on these two fields. This property can be useful if we consider the processes where the target and projectile are similar. Moreover, concerning the many loops perturbative calculations of the kernel of BK equation, it can be easier, perhaps, to use in the calculations a gauge different from the light-cone. 
As a result, the new gauge scheme can be introduced directly in the action and the many-loops calculations of the BK kernel further can be performed as defined in the paper, this task we will consider in our further publications.

 The proposed framework allows to calculate also the sub-leading corrections to the BK equation discussed in the paper. First of all, those are   corrections to the
gluon's propagator which arise from the contribution to it from the classical transverse field. From the Lipatov's formulation of the high energy scattering we know that the action can be written in terms of the physical degrees of freedom which are longitudinal reggeized gluons. Therefore, our local in rapidity  action can be fully written in terms of the Reggeon, i.e. the action can reformulated as RFT Regge Field Theory (RFT) action. Practically it means, that the transverse fields are determined in terms of the longitudinal ones, we have only two degrees of freedom after all. The 
gluon's propagator in this case gains a correction from the transverse field, which, in turn, appears as correction from the longitudinal $\B_{-}$ reggeon field, see \eq{trn14}-\eq{trn16}. 
This correction is definitely sub-leading at the shock wave regime and in the case of a non-symmetrical scattering, i.e. when the $\B_{+}$ field is enhanced in comparison to $\B_{-}$ 
due their coupling to the different impact-factors. Nevertheless, the correction can be important when the projectile and target are symmetrical or if we want to account the corrections to BK beyond shock wave approximation, this is an additional interesting issue for the further investigation of the approach. An another correction we discussed is the quark's spin contribution to the BK equation, we consider the 
processes without helicity flip, similar and more general description of the similar processes can be found in \cite{KovFl} for example. We note, that in general this process is incorporated in the framework,
but of course, when we will need to consider the helicity flip of the quark, then we will need to redefine the quark's propagator in accordance with the Section \ref{SS} prescription.
In summary, we consider the proposed approach a very promising  framework for many other calculations in the BFKL and BK physics of high energy scattering.

 The authors are indebted to Ian Balitsky for the fruitful discussions and  the helpful advises in the course of the preparation of the current paper.
 

\end{document}